\documentstyle[prd,aps,preprint,eqsecnum,epsfig,floats]{revtex}
\def\bege{\begin{equation}}
\def\ende{\end{equation}}

\def\beq{\begin{equation}}
\def\eeq{\end{equation}}
\def\bea{\begin{eqnarray}}
\def\eea{\end{eqnarray}}

\begin{document}
\tighten

\title{Disentangling Intertwined Embedded-States and Spin Effects
in Light-Front Quantization}

\author{Bernard L. G. Bakker$^1$ and Chueng-Ryong Ji$^2$}

\address{$^1$ Department of Physics and Astrophysics, Vrije Universiteit, De 
Boelelaan 1081, NL-1081 HV Amsterdam, Netherland}
\address{$^2$ Department of Physics, North Carolina State University,
Raleigh, NC 27695-8202, USA}


\maketitle

\begin{abstract}

Despite the common belief, it is not always guaranteed that
the light-front energy integration of the covariant 
Feynman amplitude automatically generates the equivalent
amplitude in light-front quantization. Our example of a light-front
calculation with a fermion loop explicitly shows that the persistent
end-point singularity in the nonvalence contribution to the bad 
component of the current,
$J^-$, leads to an infinitely different result from that obtained by
the covariant Feynman calculation unless the divergence is properly
subtracted. Ensuring the equivalence to the Feynman amplitude, we have
identified the divergent term that needs to be removed from $J^-$.
Only after this term is subtracted, the result is covariant
and satisfies current conservation. The same calculation with the 
boson loop, however, doesn't exhibit such a singular behavior and without
any adjustment yields the result identical to the Feynman amplitude.
Numerical estimates of the nonvalence contributions are presented
both for the cases of fermion and boson constituents.

\end{abstract}
\pacs{ }

\section{Introduction}
\label{sec.1}

With the advent of Light-Front field theory, one can be quite hopeful
to develop a connection between Quantum Chromodynamics (QCD) and
the relativistic constituent quark model which is used in various
electroweak form factor calculations.
QCD provides a fundamental description of
hadronic and nuclear structure in terms of elementary quark and gluon
degrees of freedom.  It is very succesful in the perturbative regime,
for instance when it is used in the explanation of the evolution of
distribution functions in deep-inelastic scattering. There, the basic
hard interaction is described by perturbative QCD and the Electro-weak
interaction. The soft parts, distribution and fragmentation functions,
are  non-perturbative ingredients, which are less well understood.
Lattice calculations are becoming increasingly accurate, but there is
still much room and need for other non-perturbative approaches.

It is part of the nature of the description of hadronic systems
in terms of quarks and gluons that the characteristic momenta
are of the same order or even very much larger than the masses of the
particles involved. Therefore a relativistic treatment is called for. A
very promising technique is Light-Front Dynamics (LFD), which treats
relativistic many-body effects in a consistent way \cite{BPP}.
In LFD a Fock-space expansion of bound states is made. The wave
function $\psi_n(x_i, k^\perp_i, \lambda_i)$ describes the
component with $n$ constituents, with longitudinal momentum fraction
$x_i$, perpendicular momentum $k^\perp_i$ and helicity $\lambda_i$,
$i=1, \dots, n$. It is the aim of LFD to determine those wave functions
and use them in conjunction with hard scattering amplitudes to describe
the properties of hadrons and their response to electroweak probes.

Recently, important steps were taken towards a realization of
this goal. In the work of Brodsky,Hiller and McCartor \cite{Hil},
it is demonstrated how to solve the problem of renormalizing light-front
Hamiltonian theories while maintaining Lorentz symmetry and other 
symmetries.
(The genesis of the work presented in \cite{Hil} may be found in 
\cite{RM} and additional examples including the use of LFD methods to 
solve the bound-state problems in field theory can be found in the review 
\cite{BPP}).
These results are indicative of the great potential of LFD for a fundamental
description of non-perturbative effects in QCD. However, at present
there are no realistic results available for wave functions of hadrons
based on QCD alone. In order to calculate the response of hadrons to external
probes, one might resort to the use of model wave functions.
This way to estimate matrix elements was used by Ji et al.\cite{Ji}.
The same reasons that make LFD so attractive to solve
bound-state problems in field theory make it also useful for a
relativistic description of nuclear systems. Presently, it is realized
that a parametrization of nuclear reactions in terms of
non-relativistic wave functions must fail.  LF methods have the
advantage that they are formally similar to time-ordered many-body
theories, yet provide relativistically invariant observables.

\begin{figure}[ht]
\begin{center}
\epsfig{figure=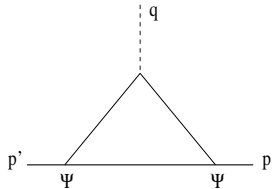,height=25mm,width=35mm}
\caption{Covariant triangle diagram \label{fig1}}
\end{center}
\end{figure}

Until now we have sketched a rather rosy picture for the
application of LFD to hadron physics. However, not all is well
and this is just the reason for the present investigation. Since the 1980's
it has been assumed that the observables computed in the framework of
LFD using the methods of perturbation theory are invariants, just as in
covariant perturbation theory. Many authors have shown that LFD has this 
feature in particular cases and some years ago some general statements to the
same effect could be made \cite{LB,SB}. A case in point is the calculation
of a current matrix element in quantum field theory.
A typical amplitude is given by the triangle diagram. One encounters this
diagram e.g.  when computing the pion form factor (see Fig.~\ref{fig1}).

\begin{figure}[ht]
\begin{center}
 \epsfig{figure=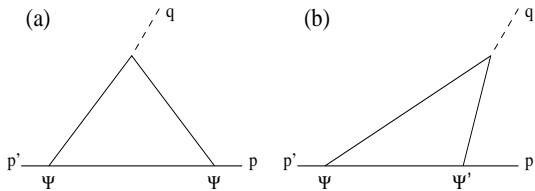,height=25mm,width=70mm}
 \caption{Time-ordered triangle diagrams \label{fig2}}
\end{center}
\end{figure}

The vertices denoted by $\Psi$ are coupling constants in covariant
perturbation theory. The hard scattering process is the absorption of a
photon of momentum $q$ by a (anti-)quark. In the LFD approach the
covariant amplitude is replaced by a series of LF time-ordered
diagrams. In the case of the triangle diagram they are depicted in
Fig.~\ref{fig2}.
The first of these two diagrams is easily interpreted in terms of the
LF wave functions $\Psi$. However, the other diagram has a vertex that
can again be written in the same way as before, but it contains also
another vertex, denoted by $\Psi'$, that cannot be written as a LF
wave function. It is a new element in LFD. We call this vertex the
{\em non-wave-function vertex}. In order to compute the form factors in the
time-like region, the contributions from these vertices must
be included. Semileptonic meson decay processes also require the
contributions from these vertices. One may try to avoid using them by
choosing special kinematic conditions. It is known however that this
will not be a simple task \cite{Fred,Jaus}.

In the present work, we investigate the contributions from the
non-wave-function vertices.
We construct both the wave-function and non-wave-function vertices using
pointlike covariant ones. 
The model used here is essentially an extension of Mankiewicz
and Sawicki's $(1+1)$-dimensional quantum field theory model \cite{SM}, 
which was later reinvestigated by several others \cite{SB,GS,Sawicki,CJ,BH}.
The starting model wave function is the solution of the covariant
Bethe-Salpeter equation in the ladder approximation with a
relativistic version of the contact interaction \cite{GS}.
The covariant model wave function is a product of two free single 
particle propagators, the overall
momentum-conserving Dirac delta function, and a constant vertex function.
Consequently, all our form factor calculations are 
various ways of evaluating the Feynman triangle
diagram in quantum field theory.

The importance of the contributions of the non-wave-function vertices was 
investigated in two cases: the electromagnetic form factors of a scalar 
and a pseudoscalar meson with spin-1/2 constituents. We also calculated 
the same form factor of a scalar meson with spin-0 constituents to see 
the spin effects. 
In 3+1 dimensions both the covariant and the LF calculations are divergent
and the model without any smeared vertex for the fermion loop is not well 
defined. This is in dramatic contrast to the case of spin-0 (boson)
constituents, where regularization is not needed at all.
In order to disentangle the issue of the
non-wave-function vertices from the need of regularization, we performed
our calculations in 1+1 dimensions, where at least the covariant
calculations for spin-1/2 constituents give finite results.

It is commonly believed and widely used that the LF energy
integration of the covariant Feynman amplitude generates the corresponding
equivalent amplitude in the LFD. As we will show in this work, however,
the equivalence between the LFD and the covariant Feynman calculation is 
not always guaranteed. The bad component of the current, $J^-$, with 
spin-1/2 constituents
exhibits a persistent end-point singularity in the contribution from 
the non-wave-function vertex. Unless the divergence in this contribution
is properly subtracted, the singular behavior leads to an infinitely
different result from that obtained by the covariant Feynman calculation.
Ensuring the equivalence to the Feynman amplitude, we have
identified the divergent term that needs to be removed from $J^-$.
Only after the identified term is subtracted, the result is covariant
and satisfies the current conservation.

In the next Section (Section~\ref{sec.2}), we present both the covariant Feynman
calculations and the LF calculations using the LF energy integration
for the electromagnetic form factors of a pseudoscalar and a scalar meson
with spin-1/2 constituents as well as the same form factor of a scalar 
meson with spin-0 constituents. Section~\ref{sec.6} contains the numerical estimates
of both the wave-function and non-wave-function vertices to the 
electromagnetic form factors of each case presented in Section~\ref{sec.2}.
The conclusion and discussion follow in Section~\ref{sec.7}.

\section{Calculations}
\label{sec.2}

The electromagnetic form factors can be extracted from the matrix
elements of the current $J^\mu$
\begin{equation}
\langle p' | J^\mu | p \rangle = i e_m (p^{\prime\mu} + p^\mu) F(q^2),
 \label{eq.01}
\end{equation}
where $e_m$ is the charge of the meson and $q^2 = (p'-p)^2$ is the
square of the four momentum transfer. If one uses the plus-component,
$J^+ = (J^0 + J^3)/\surd 2$, the LF calculation gives two finite
contributions, the wave-function part and the non-wave-function part,
that add up to the covariant result, as expected.  The importance of
the non-wave-function contribution varies strongly with the momentum
transfer and depends sensitively on the binding energy of the meson.
For small values of $q^2$ and small binding energy, the wave-function
part is dominant, but elsewhere the non-wave-function is essential for
agreement between the LF calculation and the covariant results.

The form factor can also be extracted from the minus-component of the
current, $J^- = (J^0 - J^3)/\surd 2$. Covariance guarantees that it makes
no difference whether the form factor is determined using the plus or the
minus current matrix element. As LFD is not manifestly covariant, it may
happen that $J^-$ leads to a form factor different from the one determined
using $J^+$.  As we show in this Section, the matrix element of $J^-$ 
diverges in LFD. Unless one regulates $J^-$, the current cannot be 
conserved. To assure the current conservation, it is crucial to identify 
the term that causes the divergence.  
We have identified this term exactly and found that it is
an infinite function of the momentum transfer.
If this infinite term is
subtracted, the two LF contributions become finite
as it must be in the conserved current.
Moreover, their sum equals again the covaraint result as expected.
However, the regularized LF
contributions are different from the two parts of the form factor
extracted from the plus current. The differences grow with increasing
binding energy.

\subsection{ Pseudoscalar Meson with the Fermion Loop}
\label{sec.3}

The covariant fermion triangle-loop (Fig.~\ref{fig1}) for the pseudoscalar 
meson leads to the amplitude given by
\begin{equation}
\langle p' | J^\mu | p \rangle = 4 N \int \frac{d^2 k}{(2\pi)^2}
\frac{(m^2-k^2+p\cdot p^\prime) k^\mu + (k^2 - m^2 - k\cdot p^\prime)p^\mu
+(k^2-m^2-k\cdot p) p^{\prime\mu}}{(k^2-m^2+i\epsilon)((k-p)^2-m^2+i\epsilon)
((k-p^\prime)^2-m^2+i\epsilon)},
 \label{eq.02}
\end{equation}
where $m$ is the fermion mass and $N$ modulo the obvious charge factor 
$e_m$ is the normalization 
constant fixed by the unity of the form factor at zero momentum transfer.
Even though we will present the unequal constituent mass case 
such as the kaon in the next Section of our numerical analysis, 
for the clarity of presentation we will focus in this Section on the equal mass 
case, such as the pion, only.

The usual Feynman parametrization and the covariant integration yields
\begin{equation}
 \langle p' | J^\mu | p \rangle = i(p^\mu +p^{\prime\mu})\frac{N}{\pi} 
 {\int_0}^1 {dx}{\int_0}^{1-x}{dy}\frac{2m^2(1-x)-(m^2+(x+y-1)^2M^2 -
 xyq^2)x}{((x+y)(x+y-1)M^2+m^2-xyq^2)^2},
 \label{eq.03}
\end{equation}
where $M$ is the meson mass.
For $q^2=0$, the integration leads to fix the normalization as
\begin{equation}
1/N = \frac{4m^2}{\pi M(4m^2-M^2)} \left[ \frac{M}{4m^2} + 
 \frac{1}{\sqrt{4m^2-M^2}}\arctan\left(\frac{M}{\sqrt{4m^2-M^2}}\right) \right].
 \label{eq.04}
\end{equation}

In LFD, the form factor $F(q^2)$ can be obtained by calculating
either $\langle p' | J^+ | p \rangle$ or $\langle p' | J^- | p \rangle$.
In principle, the result must be identical to the above covariant Feynman 
result regardless of which component of the current is used.
However, this is not necessarily the case as we demonstrate in the following.

First, the calculation of $\langle p' | J^+ | p \rangle$ 
integrating out the LF energy $k^-$ in Eq.~(\ref{eq.02}) yields $F(q^2)$ given by
\begin{eqnarray}
F(q^2) &=& \frac{N}{\pi(2+\alpha)} \Biggl{[}
{\int_0}^1 {dx} \frac{(1+\alpha)^2m^2}{[m^2-x(1-x)M^2][(1+\alpha)^2m^2-
(1-x)(\alpha+x)M^2]} \; 
\nonumber 
\\ 
&+& {\int_0}^\alpha {dx} \frac{(1+\alpha)(\alpha-x)[x(\alpha-x)M^2-
(1+\alpha)^2m^2]}{\alpha [(\alpha-x)(1+x)M^2-(1+\alpha)^2m^2]
[x(\alpha-x)M^2 + (1+\alpha)m^2]} \Biggl{]},
 \label{eq.05}
\end{eqnarray}
where $\alpha$ is given by $q^2 = - \frac{\alpha^2 
M^2}{1+\alpha}$. In Eq.~(\ref{eq.05})
the first and second terms correspond to the contributions from
the wave-function and non-wave-function vertices depicted in the first
and second diagrams in Fig.~\ref{fig2}, respectively.
We have verified that the contribution from the non-wave-function part
vanishes at $\alpha = 0$, {\it i.e.} at $q^2 = 0$, indicating the absence
of a zero-mode contribution \cite{CJ} in the good component of the current
$J^+$. Adding both contributions in Eq.~(\ref{eq.05}), we obtain
\begin{eqnarray}
F(q^2) &=& \frac{2N(1+\alpha)m^2}{\pi \alpha M 
[(2+\alpha)^2m^2-(1+\alpha)M^2]} \nonumber \\
 & &  \left[ \frac{2+2\alpha+\alpha^2}
 {\sqrt{4(1+\alpha)m^2+\alpha^2 M^2}} {\rm Artanh} \left( \frac{\alpha M}
 {\sqrt{4(1+\alpha)m^2+\alpha^2 M^2}}\right) \right. \nonumber \\
 && \left. \rule{40mm}{0mm}
 + \frac{2\alpha}{\sqrt{4m^2-M^2}}\arctan\left(\frac{M}{\sqrt{4m^2-M^2}} \right)
\right] .
 \label{eq.06}
\end{eqnarray}
This result is identical to the form factor obtained by the covariant
Feynman calculation given by Eq.~(\ref{eq.03}) and F(0)=1 gives the
same normalization given in Eq.~(\ref{eq.04}).

On the other hand, the calculation of $\langle p' | J^- | p \rangle$ 
yields $F(q^2)$ given by
\begin{eqnarray}
F(q^2) &=& \frac{N}{\pi(2+\alpha)} \Biggl{[} (1+\alpha) M^2
{\int_0}^1 {dx} \frac{(1-x)^2}{[m^2-x(1-x)M^2][(1+\alpha)^2m^2-
(1-x)(\alpha+x)M^2]} \; 
\nonumber 
\\ 
&-& \frac{(1+\alpha)^2m^2}{\alpha M^2} {\int_0}^\alpha {\frac{dx}{\alpha-x}} 
\frac{(1+\alpha)^2 m^2 + 
\{(1+\alpha)^2-(1+x)\}(\alpha-x)M^2}
{[(\alpha-x)(1+x)M^2-(1+\alpha)^2m^2]
[x(\alpha-x)M^2 + (1+\alpha)m^2]} \Biggl{]},
 \label{eq.07}
\end{eqnarray}
where again it is apparent that
the first and second terms correspond to the contributions from
the wave-function and non-wave-function vertices, respectively.
However, the non-wave-function part shows the end-point singularity
coming from $1\over \alpha-x$.
Without subtracting the end-point singularity, the result
is infinitely different from that obtained in the $\langle p' | J^+ | p 
\rangle$ calculation. This is an astonishing result that deviates
from the common belief in the equivalence of the LFD and the covariant 
Feynman calculation. Neither covariance nor current conservation 
is satisfied without a certain adjustment. In order to identify the term 
that must be subtracted, we rewrite the above equation as follows:
\begin{eqnarray}
F(q^2) &=& \frac{N}{\pi(2+\alpha)} \Biggl{[} (1+\alpha) M^2
{\int_0}^1 {dx} \frac{(1-x)^2}{[m^2-x(1-x)M^2][(1+\alpha)^2m^2-
(1-x)(\alpha+x)M^2]} \; 
\nonumber 
\\ 
 & &\rule{34mm}{0mm}+{\int_0}^\alpha {dx}\frac{R(x,\alpha)}{\alpha-x} \Biggl{]},
 \label{eq.08}
\end{eqnarray}
where $R(x,\alpha)$ is defined by
\begin{equation}
R(x,\alpha) = - \frac{(1+\alpha)^2m^2[(1+\alpha)^2m^2+\{(1+\alpha)^2-(1+x)\}
(\alpha-x)M^2}{\alpha M^2
[(\alpha-x)(1+x)M^2-(1+\alpha)^2m^2]
[x(\alpha-x)M^2 + (1+\alpha)m^2]}.
 \label{eq.09}
\end{equation}
In order to obtain the identical result to Eq.~(\ref{eq.06}) from 
the $\langle p' | J^+ | p \rangle$ calculation,
we find that $R(\alpha,\alpha)$ must be subtracted from the numerator
of the non-wave-function part integrand, {\it i.e.},
\begin{eqnarray}
F(q^2) &=& \frac{N}{\pi(2+\alpha)} \Biggl{[} (1+\alpha) M^2
{\int_0}^1 {dx} \frac{(1-x)^2}{[m^2-x(1-x)M^2][(1+\alpha)^2m^2-
(1-x)(\alpha+x)M^2]} \; 
\nonumber \\ 
 & & \rule{34mm}{0mm} + {\int_0}^\alpha {dx} \frac{R(x,\alpha)-R(\alpha,\alpha)}
{\alpha-x}\Biggl{]}.
 \label{eq.10}
\end{eqnarray}
The subtracted term $R(\alpha,\alpha) = \frac{1+\alpha}{\alpha M^2}$
depends on the momentum transfer and never vanishes. 
While the subtracted result Eq.~(\ref{eq.10}) is identical to Eq.~(\ref{eq.06}),
it is interesting to note that the zero-mode contribution
does not vanish in the $\langle p' | J^- | p \rangle$ calculation
as the non-wave-function part in Eq.~(\ref{eq.10}) still survives even at $q^2=0$. 
However, the subtracted result Eq.~(\ref{eq.10}) with the zero-mode 
contribution assures covariance and satisfies current conservation.

\subsection{ Scalar Meson with the Fermion Loop}
\label{sec.4}

The Feynman parametrization and the covariant integration 
of the fermion triangle-loop for the scalar meson gives the 
following amplitude:
\begin{eqnarray}
\langle p' | J^\mu | p \rangle &=& 4 N \int \frac{d^2 k}{(2\pi)^2}
\frac{(3m^2+k^2-p\cdot p^\prime) k^\mu - (k^2 + m^2 - k\cdot p^\prime)p^\mu
-(k^2+m^2-k\cdot p) p^{\prime\mu}}{(k^2-m^2+i\epsilon)((k-p)^2-m^2+i\epsilon)
((k-p^\prime)^2-m^2+i\epsilon)} \;
\nonumber \\
&=& i(p^\mu +p^{\prime\mu})\frac{N}{\pi} 
{\int_0}^1 {dx}{\int_0}^{1-x}{dy}\frac{x((1-x-y)^2M^2-m^2-xyq^2)}
{((x+y)(x+y-1)M^2+m^2-xyq^2)^2},
 \label{eq.11}
\end{eqnarray}
where the normalization $N$ is again
fixed by $F(0)=1$ and given by
\begin{equation}
1/N = \frac{4m^2}{\pi M^3} \left[ \frac{M}{4m^2} - \frac{1}
{\sqrt{4m^2-M^2}} \arctan \left( \frac{M}{\sqrt{4m^2-M^2}} \right) \right].
 \label{eq.12}
\end{equation}

In LFD, the calculation of $\langle p' | J^+ | p \rangle$ 
leads to $F(q^2)$ given by
\begin{eqnarray}
F(q^2) &=& \frac{N}{\pi(2+\alpha)} \Biggl{[}
{\int_0}^1 {dx} 
\frac{(1+\alpha)\{2(1-x)(2x+\alpha)-(1+\alpha)\}m^2}
{[m^2-x(1-x)M^2][(1+\alpha)^2m^2-(1-x)(\alpha+x)M^2]} \; 
\nonumber 
\\ 
&+& {\int_0}^\alpha {dx} \frac{(1+\alpha)(\alpha-x)[(1+\alpha)
(1+4x-\alpha)m^2-x(\alpha-x)M^2]}
{\alpha [(\alpha-x)(1+x)M^2-(1+\alpha)^2m^2]
[x(\alpha-x)M^2 + (1+\alpha)m^2]} \Biggl{]},
 \label{eq.13}
\end{eqnarray}
where the contribution from the non-wave-function part again
vanishes at $q^2 = 0$, indicating the absence
of zero-mode contribution \cite{CJ} in the good component of the current,
$J^+$. Adding both contributions, we find
\begin{eqnarray}
F(q^2) &=& \frac{2N(1+\alpha)m^2}{\pi \alpha M^3 
[(2+\alpha)^2m^2-(1+\alpha)M^2]} \;
\nonumber \\
&&\Biggl{[}\frac{8(1+\alpha)m^2
-(2+2\alpha-\alpha^2)M^2}{\sqrt{4(1+\alpha)m^2+\alpha^2 M^2}} 
{\rm Artanh} \left( \frac{\alpha M}
{\sqrt{4(1+\alpha)m^2+\alpha^2 M^2}} \right)
\nonumber \\
 & & \rule{50mm}{0mm} - \frac{2\alpha}{\sqrt{4m^2-M^2}}
 \arctan \left( \frac{M}{\sqrt{4m^2-M^2}} \right)
\Biggl{]} .
 \label{eq.14}
\end{eqnarray}
This result is identical to the form factor obtained by the covariant
Feynman calculation given by Eq.~(\ref{eq.11}) and F(0)=1 gives the same normalization
presented in Eq.~(\ref{eq.11}).

However, the calculation of $\langle p' | J^- | p \rangle$ 
generates an end-point point singularity similar to the one observed in
the pseudoscalar case. Defining the function
\begin{equation}
S(x,\alpha) = 
\frac{(1+\alpha)^2m^2[(1+\alpha)(1+4x-3\alpha)m^2+\{\alpha(2-\alpha)
+(2\alpha-1)x\}(\alpha-x)M^2]}
{\alpha M^2[(\alpha-x)(1+x)M^2-(1+\alpha)^2m^2]
[x(\alpha-x)M^2 + (1+\alpha)m^2]},
 \label{eq.15}
\end{equation}
we find $F(q^2)$ given by
\begin{eqnarray}
F(q^2) &=& \frac{N}{\pi(2+\alpha)} \Biggl{[} - \frac{(1+\alpha)}{M^2}
{\int_0}^1 {dx} \frac{4(1+\alpha)m^4-2(2+\alpha)(1-x)m^2M^2+(1-x)^2M^4}
{[m^2-x(1-x)M^2][(1+\alpha)^2m^2-(1-x)(\alpha+x)M^2]} \; 
\nonumber 
\\ 
&&\rule{30mm}{0mm}+{\int_0}^\alpha {dx} \frac{S(x,\alpha)}{\alpha-x} \Biggl{]},
 \label{eq.16}
\end{eqnarray}
where the non-wave-function part again shows the end-point singularity
coming from $1\over \alpha-x$.
In order to obtain the identical result to Eq.~(\ref{eq.14}) from 
the $\langle p' | J^+ | p \rangle$ calculation,
we find that $S(\alpha,\alpha)$ must be subtracted from the numerator
of the non-wave-function part integrand, {\it i.e.},
\begin{eqnarray}
F(q^2) &=& \frac{N}{\pi(2+\alpha)} \Biggl{[} - \frac{(1+\alpha)}{M^2}
{\int_0}^1 {dx} \frac{4(1+\alpha)m^4-2(2+\alpha)(1-x)m^2M^2+(1-x)^2M^4}
{[m^2-x(1-x)M^2][(1+\alpha)^2m^2-(1-x)(\alpha+x)M^2]} \; 
\nonumber \\ 
 & & \rule{30mm}{0mm}
 + {\int_0}^\alpha {dx} \frac{S(x,\alpha)-S(\alpha,\alpha)}{\alpha-x} 
\Biggl{]}.
 \label{eq.17}
\end{eqnarray}
The subtracted term $S(\alpha,\alpha) = -\frac{1+\alpha}{\alpha M^2}$
again cannot  vanish.
The zero-mode contribution is
also visible since the second term in Eq.~(\ref{eq.17}) doesn't vanish even 
if $q^2=0$. The subtracted result Eq.~(\ref{eq.17}), however, assures the 
covariance and satisfies the current conservation as in the previous
calculation of the pseudoscalar meson.
   
\subsection{ Scalar Meson with the Boson Loop}
\label{sec.5}

For a comparison with the scalar constituents neglecting spin effects,
we present in this subsection the calculation of $F(q^2)$ for a scalar
meson with a boson loop.  The Feynman parametrization and the
covariant integration of the boson triangle-loop for the scalar meson
gives the following amplitude:  
\begin{eqnarray} \langle p' | J^\mu | p
\rangle &=& N \int \frac{d^2 k}{(2\pi)^2} \frac{p^\mu + p^{\prime\mu} -
2k^\mu }
{(k^2-m^2+i\epsilon)((k-p)^2-m^2+i\epsilon)((k-p^\prime)^2-m^2+i\epsilon)} \;
\nonumber \\
&=& i(p^\mu +p^{\prime\mu})\frac{N}{4\pi} 
{\int_0}^1 {dx}{\int_0}^{1-x}{dy}\frac{2x-1}
{((x+y)(x+y-1)M^2+m^2-xyq^2)^2},
 \label{eq.18}
\end{eqnarray}
where the normalization $N$ is given by
\begin{equation}
1/N = \frac{1}{2\pi M^2 (4m^2-M^2)} \left[-1+\frac{2(2m^2-M^2)}
{M\sqrt{4m^2-M^2}}\arctan \left( \frac{M}{\sqrt{4m^2-M^2}}\right) \right].
 \label{eq.19}
\end{equation}

In LFD, the calculation of $\langle p' | J^+ | p \rangle$ 
leads to $F(q^2)$ given by
\begin{eqnarray}
F(q^2) &=& \frac{N}{4\pi(2+\alpha)} \Biggl{[} {-1}
{\int_0}^1 {dx} 
\frac{(1+\alpha)(2x+\alpha)(1-x)}
{[m^2-x(1-x)M^2][(1+\alpha)^2m^2-(1-x)(\alpha+x)M^2]} \; 
\nonumber \\ 
 & & +  {\int_0}^\alpha {dx} \frac{(1+\alpha)^2(\alpha-2x)(\alpha-x)}
{\alpha [(\alpha-x)(1+x)M^2-(1+\alpha)^2m^2]
[x(\alpha-x)M^2 + (1+\alpha)m^2]} \Biggl{]},
 \label{eq.20}
\end{eqnarray}
where the contribution from the non-wave-function part again
vanishes at $q^2 = 0$, indicating the absence
of a zero-mode contribution \cite{CJ} in the good component of the current
$J^+$ as in the fermion loop cases. Adding both contributions, we find
\begin{eqnarray}
F(q^2) &=& \frac{N(1+\alpha)}{2\pi \alpha M^3 
[(2+\alpha)^2m^2-(1+\alpha)M^2]} \;
\nonumber \\
&&\Biggl{[}-{\sqrt{4(1+\alpha)m^2+\alpha^2 M^2}} 
{\rm Artanh} \left( \frac{\alpha M}
{\sqrt{4(1+\alpha)m^2+\alpha^2 M^2}} \right)
\nonumber \\
 & & \rule{34mm}{0mm} + \frac{2\alpha(2m^2-M^2)}{\sqrt{4m^2-M^2}}
 \arctan \left( \frac{M}{\sqrt{4m^2-M^2}} \right) \Biggl{]} .
 \label{eq.21}
\end{eqnarray}
This result is identical to the form factor obtained by the covariant
Feynman calculation given by Eq.~(\ref{eq.18}) and F(0)=1 gives the same normalization
presented in Eq.~(\ref{eq.19}).

Similarly, the calculation of $\langle p' | J^- | p \rangle$ 
generates $F(q^2)$ given by
\begin{eqnarray}
F(q^2) &=& \frac{N}{4\pi(2+\alpha)} \Biggl{[} \frac{(1+\alpha)}{M^2}
{\int_0}^1 {dx}
\frac{2(1+\alpha)m^2-(2+\alpha)(1-x)M^2}
{[m^2-x(1-x)M^2][(1+\alpha)^2m^2-(1-x)(\alpha+x)M^2]} \;
\nonumber \\
 & & +\frac{(1+\alpha)^2}{\alpha M^2}{\int_0}^\alpha {dx}
\frac{2(1+\alpha)m^2+\alpha(\alpha-x)M^2}
{[(\alpha-x)(1+x)M^2-(1+\alpha)^2m^2]
[x(\alpha-x)M^2 + (1+\alpha)m^2]} \Biggl{]}.
 \label{eq.22}
\end{eqnarray}
Unlike the cases of fermion constituents, however, 
the non-wave-function part here does not exhibit the end-point singularity.
Without any adjustment, we find that Eq.~(\ref{eq.22}) is identical to
Eq.~(\ref{eq.21}) obtained from $\langle p' | J^+ | p \rangle$.
Thus, the result is automatically covariant and satisfies current
conservation. However, the zero-mode contribution is still present in
$J^-$ current as one can easily see that the second term in Eq.~(\ref{eq.22})
doesn't vanish when $\alpha$ goes to zero. In the next section,
we numerically estimate the importance of the non-wave-function part
in all three cases that we presented in this section.

\section{Numerical Results}
\label{sec.6}

We have estimated both the wave-function and non-wave-function 
vertices to the electromagnetic form factors of each case presented in 
Section~\ref{sec.2}. For the three cases (A,B,C) presented in Section~\ref{sec.2}, we
show the numerical results of the form factor calculated via the minus
component as well as the plus component of the current.
We denote the contributions from the wave-function and non-wave-function 
vertices as $F_{\rm val}^{+(-)}$ and $F_{\rm nv}^{+(-)}$, respectively when the
plus(minus) component of the current is used.
The sum of the two contributions is denoted by $F_{\rm tot}^{+(-)}$,{\it i.e.}
$F_{\rm tot}^{+(-)} = F_{\rm val}^{+(-)} + F_{\rm nv}^{+(-)}$.

For the numerical computation, we take for the experimental meson masses of
the pion and the kaon $m_\pi = 0.140$ GeV and $m_K = 0.494$ GeV resp. and 
vary the quark masses to investigate the binding-energy dependence of the
meson form factors.  We call the pseudoscalar meson with the equal quark
masses and mass $m_\pi = 0.140$ GeV the "pion".
Likewise, the pseudoscalar meson with the unequal quark masses and the
meson mass $m_K = 0.494$ GeV is called the "kaon".  

\begin{figure}
\begin{center}
\epsfig{figure=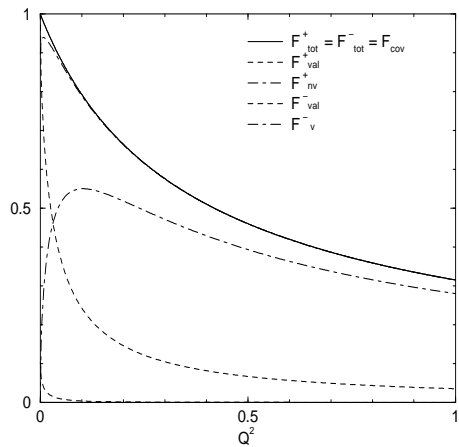,height=60mm,width=60mm}
\caption{Pion form factor in LF calculation in 1+1 Dim.
Pseudoscalar meson with spinor consitutents.
 $M = 0.140$ GeV, $m_q=0.250$ GeV. 
Fat lines correspond to the plus-current, thin lines to the minus-current.
The solid line is the full form factor. It is the sum of the valence and the
non-valence contributions. The separate contributions differ but the sums
coincide. The form factor determined from the covariant amplitude is identical
with the full form factor determined in the LF calculation.
 \label{figlfps.01}}
\end{center}
\end{figure}

The "pion" form factor with the quark mass $m_q = 0.250$ GeV is shown
in Fig.~\ref{figlfps.01}. When the plus current is used, the valence contribution $F_{\rm val}^+$
diminishes very quickly as $Q^2$ gets larger even though the normalization
at $Q^2 = 0$ is entirely coming from the valence part as we pointed out
in Section~\ref{sec.2}. The crossing between $F_{\rm val}^+$ and $F_{\rm nv}^+$ appears
at $Q^2$ below $0.05$ GeV${}^2$. When the minus current is used, however,
the valence contribution $F_{\rm val}^-$ is negligible even at $Q^2=0$ and
the entire result is essentially given by $F_{\rm nv}^-$. The value of
$F_{\rm nv}^-(0)$ corresponds to the zero-mode contribution in the minus
current $J^-$ and it is interesting to note that more than $90 \%$ of the 
form factor at $Q^2 = 0$ is contributed by the zero-mode. Since $F_{\rm tot}^+
= F_{\rm tot}^-$ exactly coincide with the covariant result obtained by Eq.(3)
for all $Q^2$ as they must, only a single solid line is depicted in 
Fig.~\ref{figlfps.01}. The same applies to all of the other figures 
presented in this work.

\begin{figure}
\hspace{17mm} (a) \hspace{46mm} (b) \hspace{46mm} (c)\\
\epsfig{figure=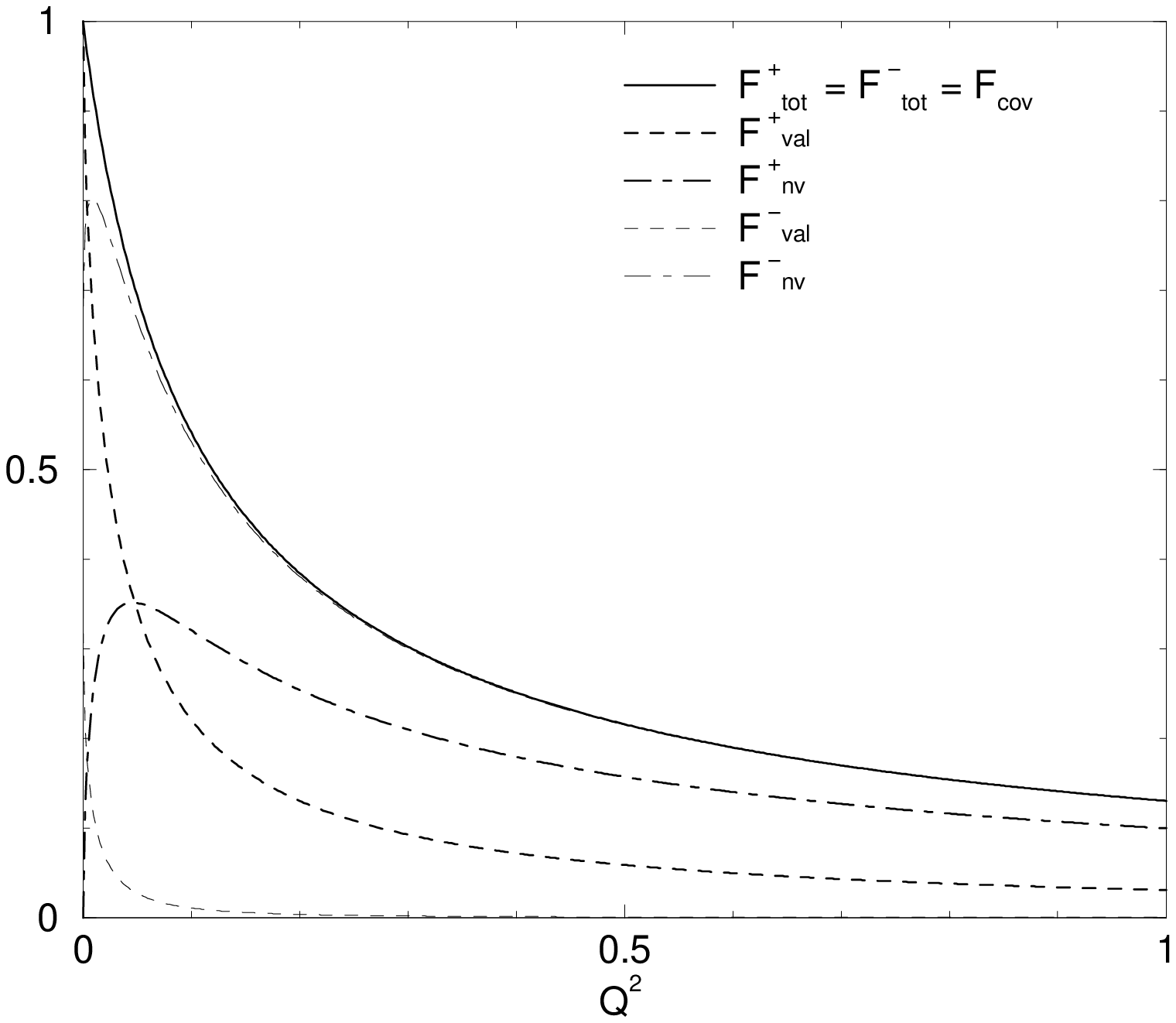,height=55mm,width=50mm} \hspace{2mm}
\epsfig{figure=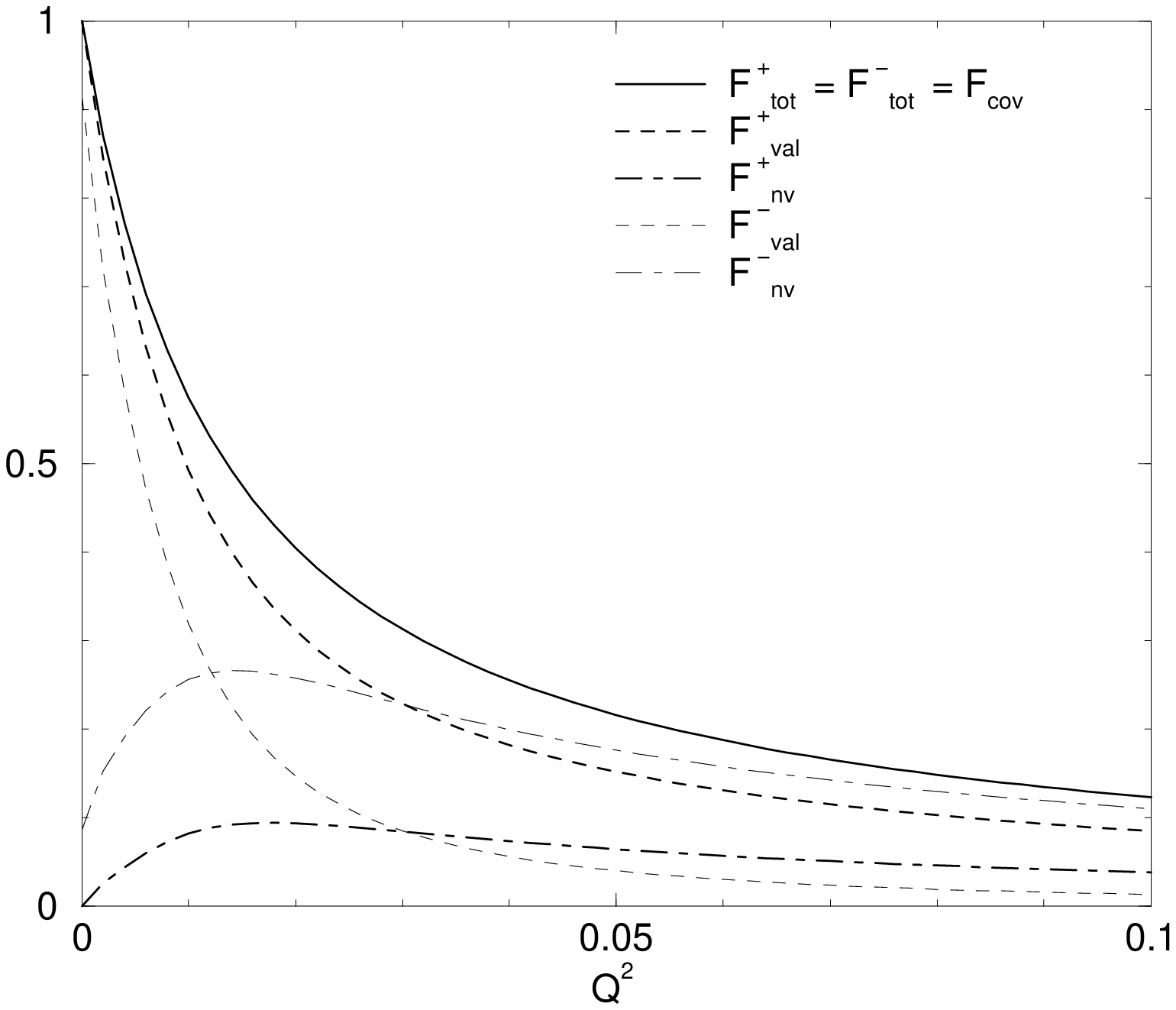,height=55mm,width=50mm} \hspace{2mm}
\epsfig{figure=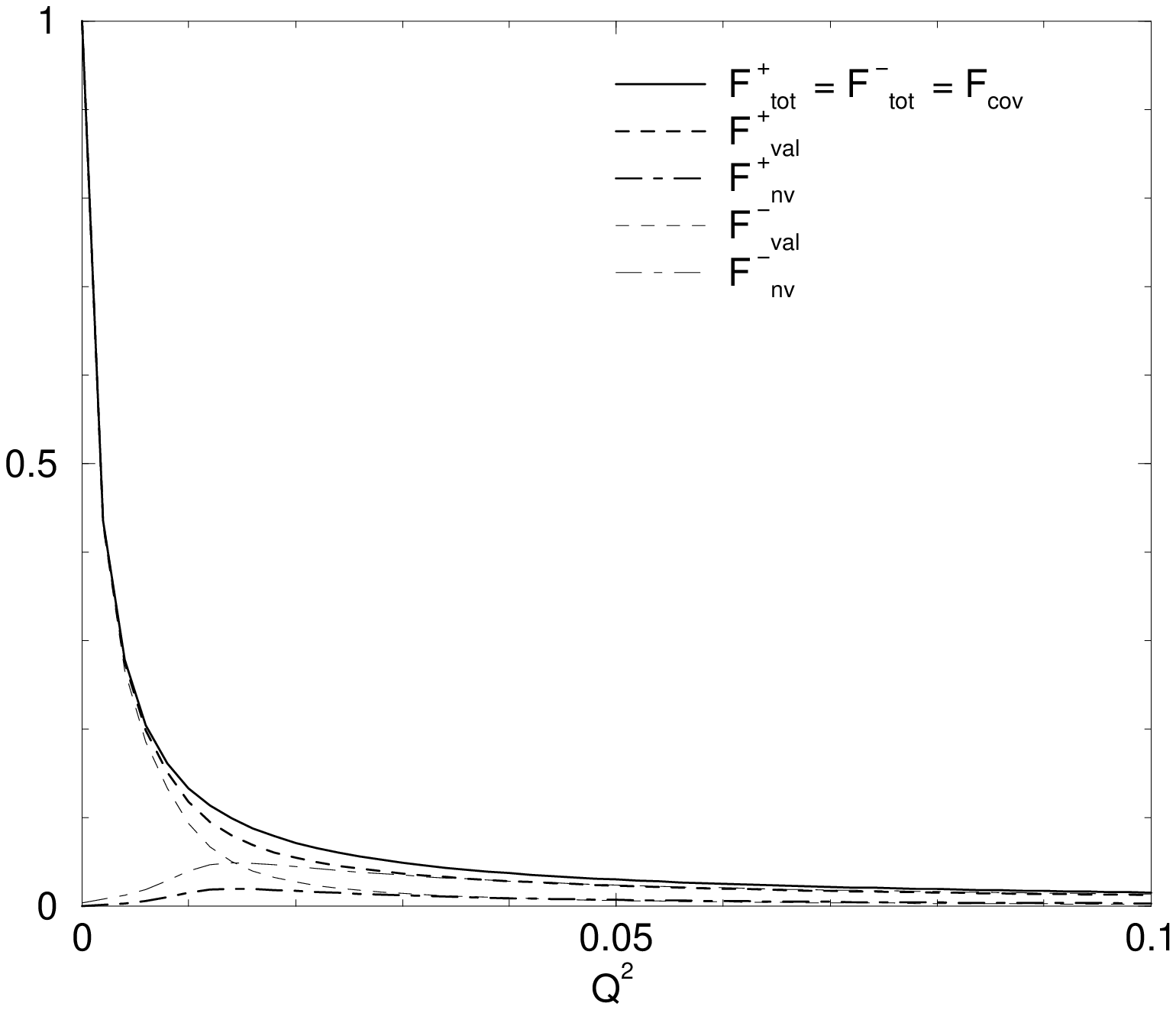,height=55mm,width=50mm}
\caption{"Pion" form factor in LF calculation in 1+1 Dim.
Pseusoscalar meson with spinor consitutents.
(a) $m_\pi = 0.140$, $m_q=0.140$. 
(b) $m_\pi = 0.140$, $m_q=0.077$. 
(c) $m_\pi = 0.140$, $m_q=0.0707$. 
The lines have the same meaning as in Fig.~\ref{figlfps.01}.
Note the change in scale in panels (b) and (c).
 \label{figlfps.04}}
\end{figure}

In Fig.~\ref{figlfps.04}, we
present the results for the "pion" by changing the quark masses  in the
following way:
$m_q = 0.140$, 0.077 and 0.0707 GeV, respectively. The closer $m_q$ is to 
$m_\pi/2 = 0.07$ GeV, the smaller the binding energy gets, and the slope of
$F_{\rm tot}^{+/-}$ at $Q^2 =0$ (or the charge radius) increases with decreasing
quark mass, as
expected. We find that the crossing between $F_{\rm val}^+$ and $F_{\rm nv}^+$
occurs at a larger value of $Q^2$ and $F_{\rm val}^-$ becomes larger near $Q^2 =
0$ as the binding gets weaker.  This may be explained by the reduction
of the probability to generate the non-wave-function vertex (or the
higher Fock state) compared to the valence state as the interaction
between the constituents gets weaker.  Thus, in the weaker binding,
$F_{\rm val}^+$ dominates over $F_{\rm nv}^+$.  Similarly, $F_{\rm val}^-$ becomes
the main contribution near $Q^2=0$ which is the only region where the
form factor exists in the weak binding limit. Consequently, the
zero-mode $F_{\rm nv}^-(0)$ gets substantially diminished as shown in
Fig.~\ref{figlfps.04}.

\begin{figure}
\hspace{17mm} (a) \hspace{46mm} (b) \hspace{46mm} (c)\\
 \hspace{-5mm}
\epsfig{figure=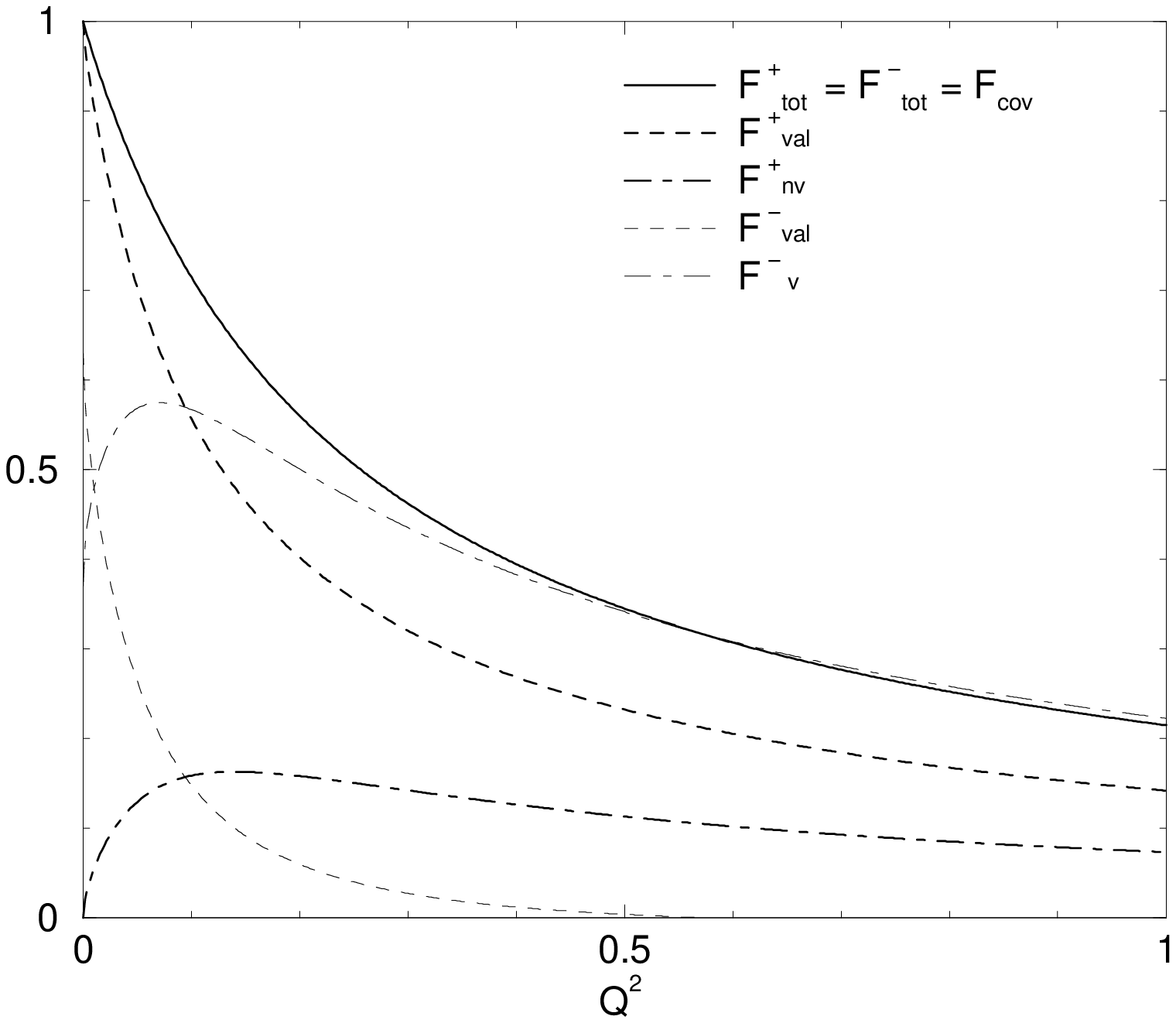,height=55mm,width=50mm} \hspace{2mm}
\epsfig{figure=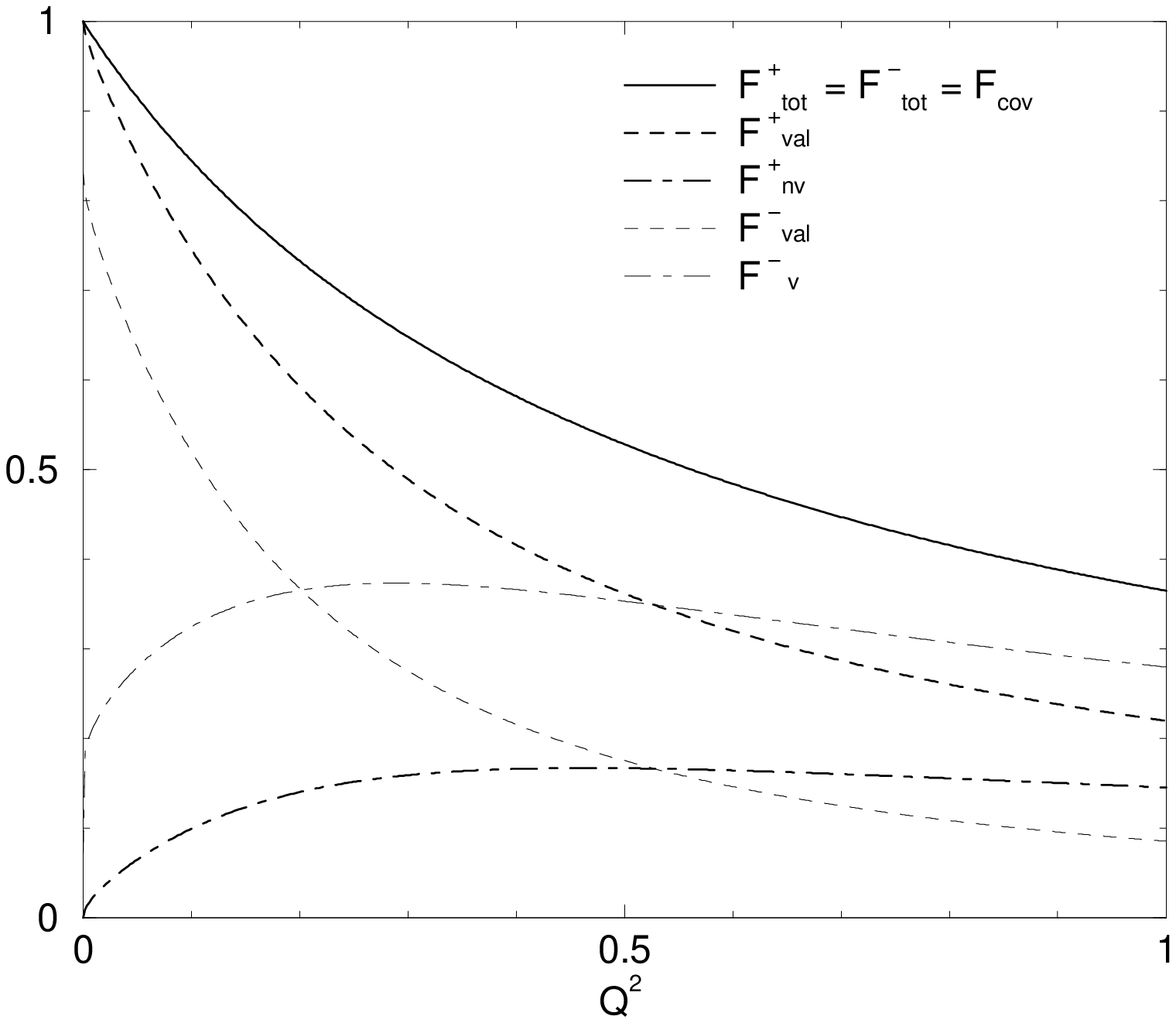,height=55mm,width=50mm} \hspace{2mm}
\epsfig{figure=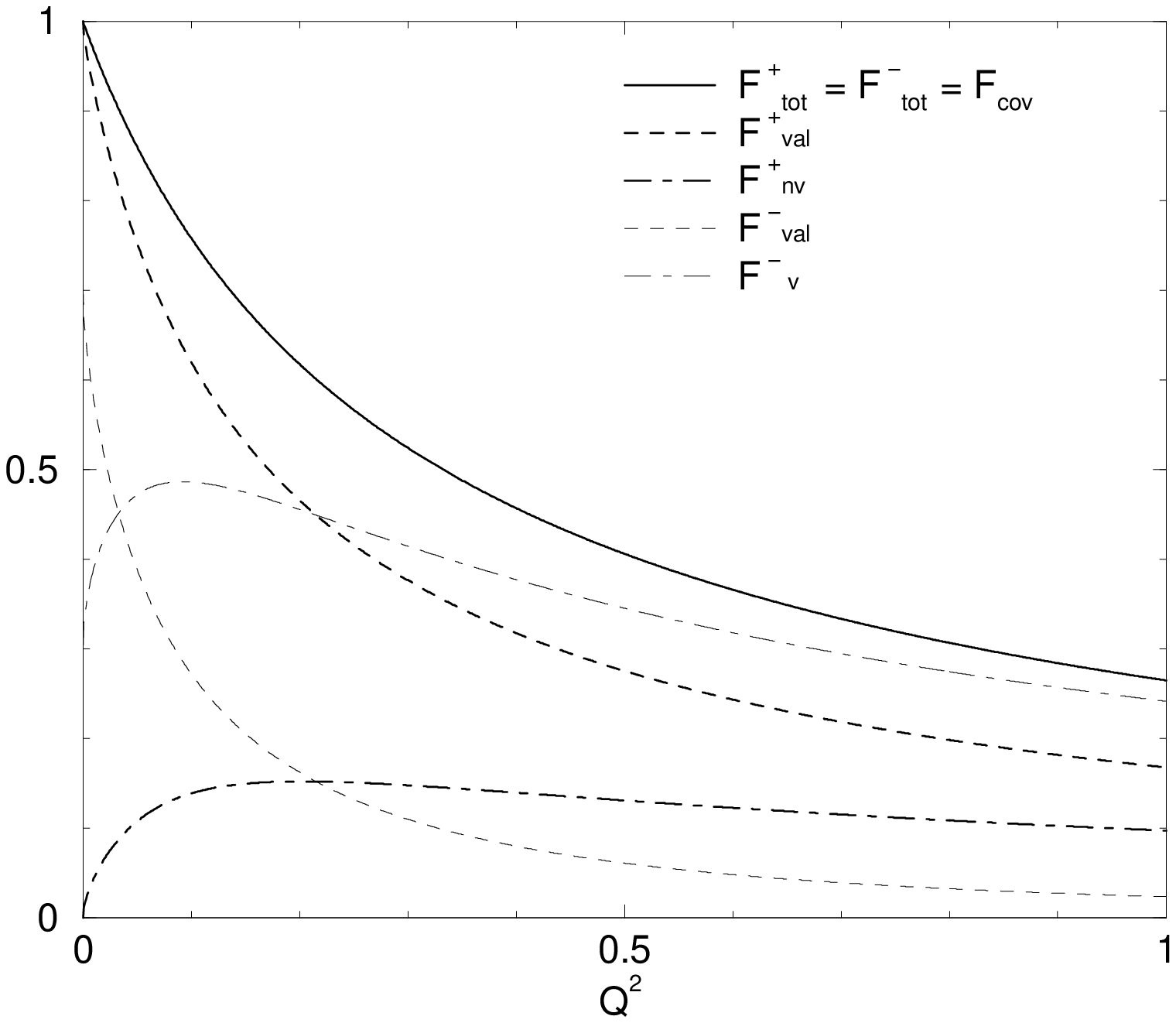,height=55mm,width=50mm}
\caption{Kaon form factor. LFD calculation in 1+1 Dim.
 $m_K = 0.494$, $m_q=0.250$, $m_s=0.370$. 
 (a) Charge $+1$ on the light quark, (b) charge $+1$ on the strange quark,
 (c) $e_q = 2/3$, $e_s = 1/2$.
The lines have the same meaning as in Fig.~\ref{figlfps.01}.
 \label{figlfps.07}}
\end{figure}

The "kaon" form factor with $m_q = 0.25$ GeV and $m_s = 0.37$ GeV is shown
in Fig.~\ref{figlfps.07}. 
Compared to the "pion" case, the dominance of $F_{\rm val}^+$ extends 
to the larger $Q^2$ region and the crossover between $F_{\rm val}^+$ and 
$F_{\rm nv}^+$ is postponed beyond the range of $Q^2 = 1$ GeV${}^2$.
The zero-mode $F_{\rm nv}^-(0)$ is also much smaller than in the "pion" case even 
though $F_{\rm nv}^-$ rises very quickly as $Q^2$ gets away from the zero 
range. As one can see in Figs.5(a) and (b), the contribution from the 
heavier quark struck by the photon is larger than that from the lighter 
quark struck by the photon. We have indeed confirmed that as $m_s$ gets 
larger only the contribution from the heavy quark struck by the photon 
dominates as expected.

\begin{figure}
\begin{center}
\hspace{3mm} (a) \hspace{59mm} (b) \\
\epsfig{figure=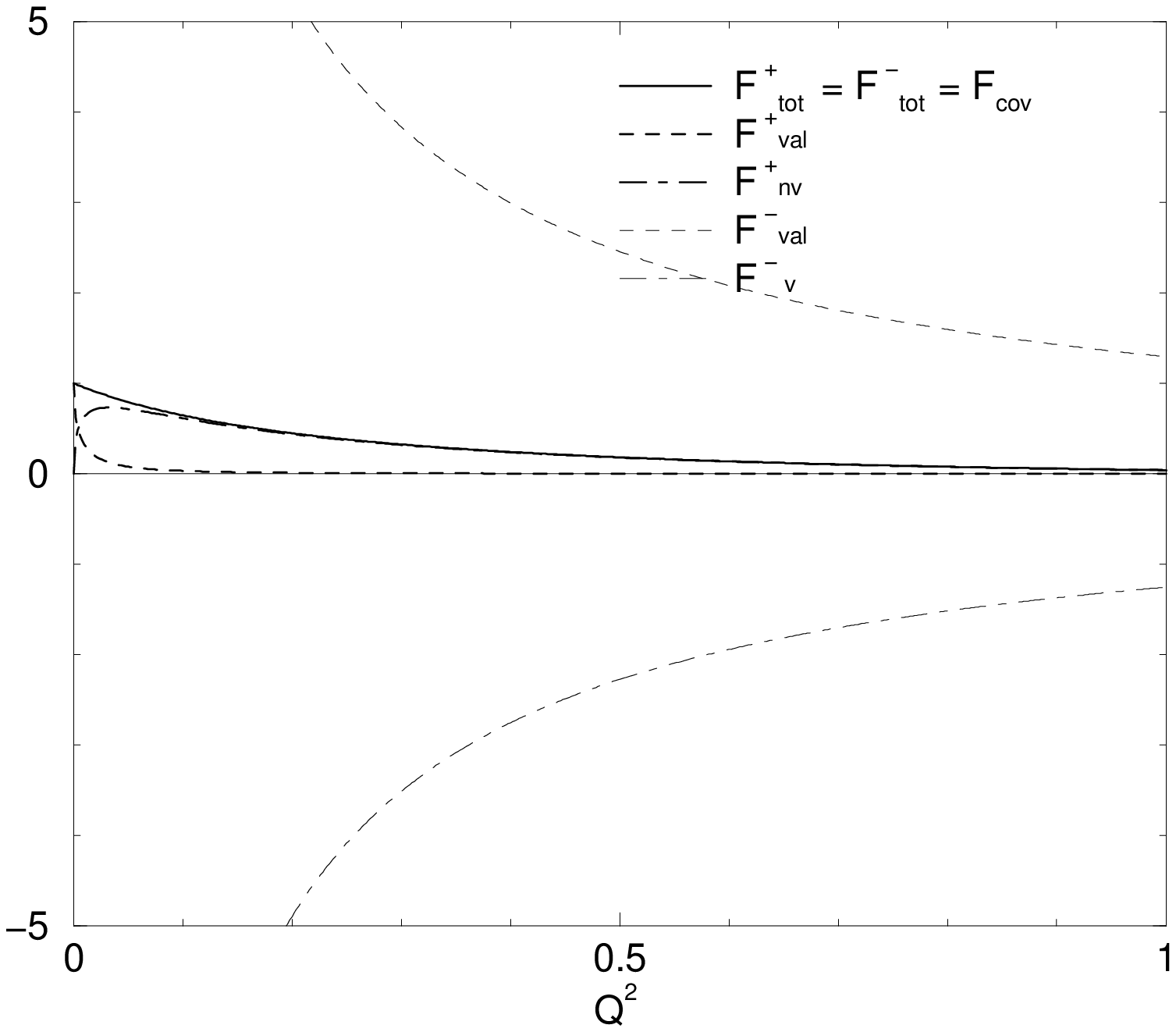,height=55mm,width=50mm} \hspace{15mm}
\epsfig{figure=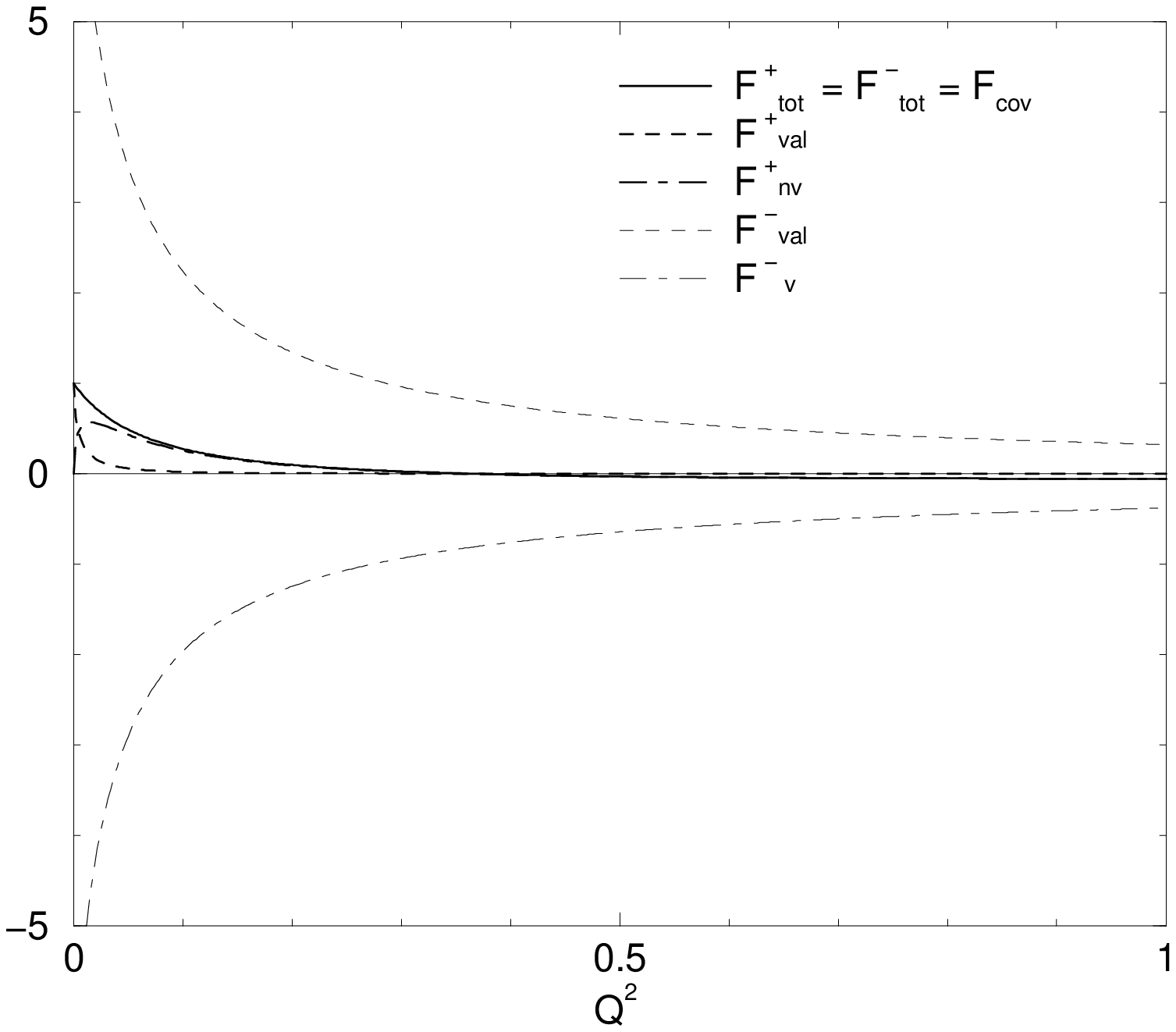,height=55mm,width=50mm} 
\end{center}
\begin{center}
\hspace{3mm} (c) \hspace{59mm} (d) \\
\epsfig{figure=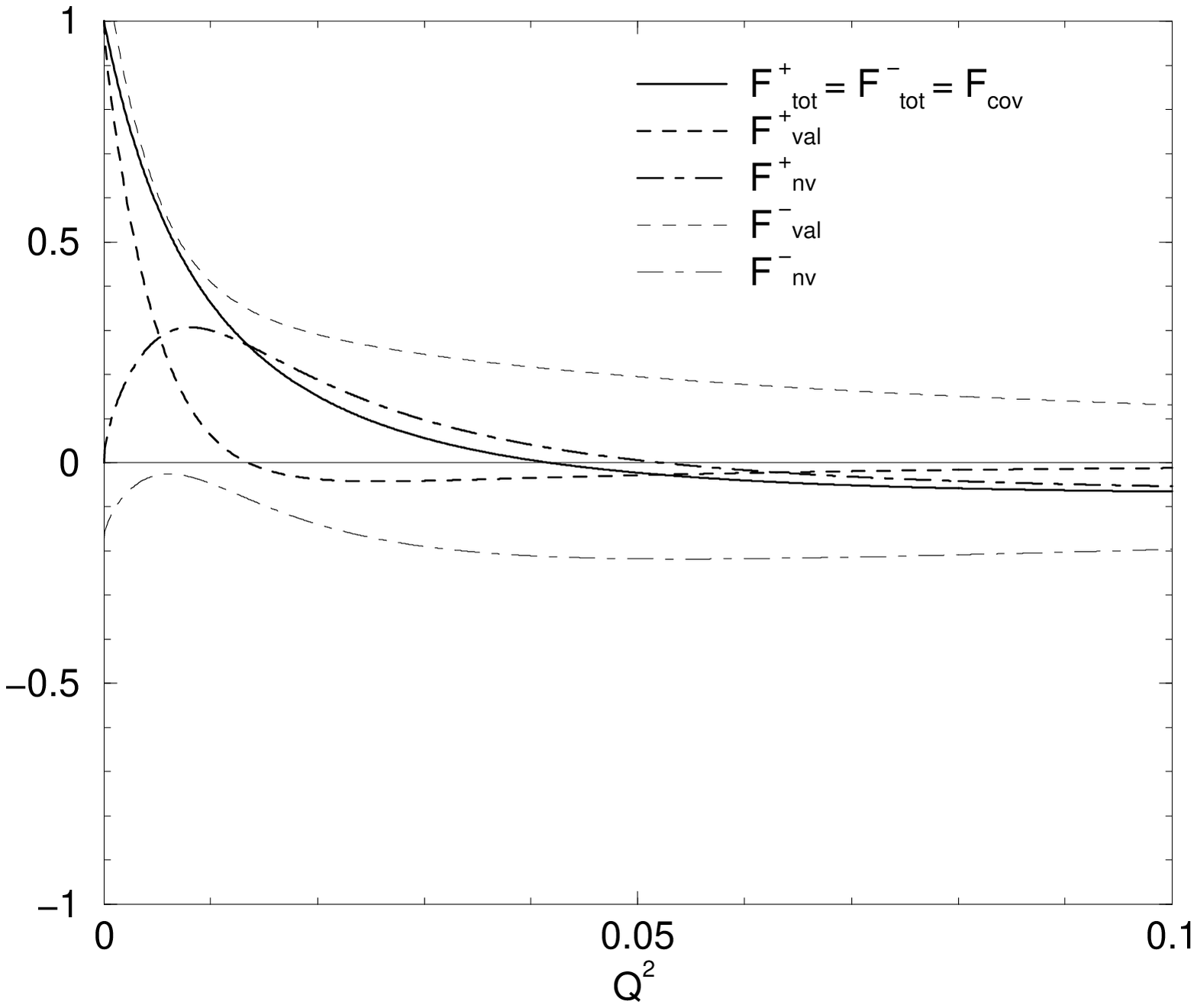,height=55mm,width=50mm} \hspace{15mm}
\epsfig{figure=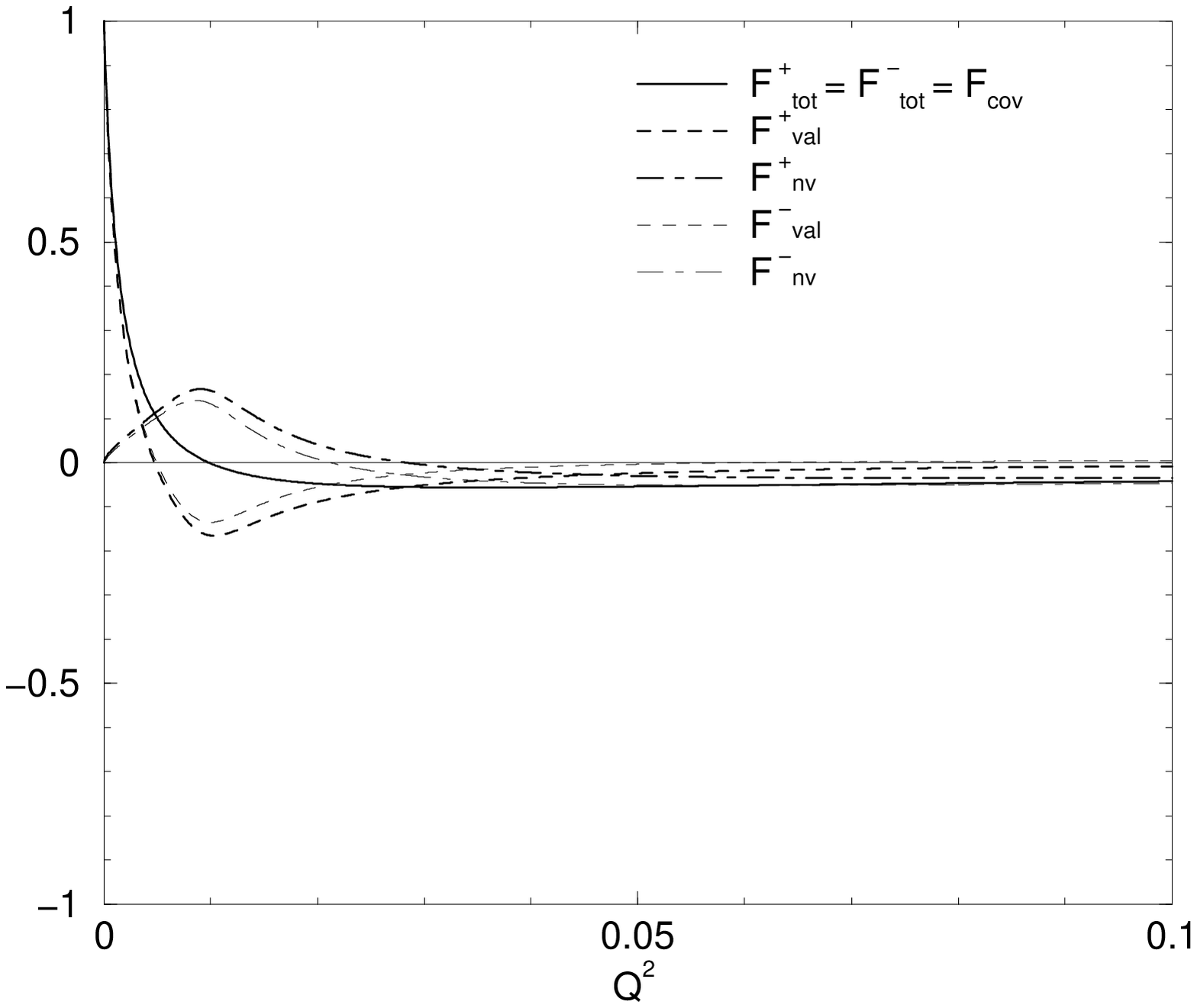,height=55mm,width=50mm}
\caption{"Pion" form factor in LF calculation in 1+1 Dim.
 Scalar meson with spinor constituents.
 (a) $m_\pi = 0.140$, $m_q=0.250$. 
 (b) $m_\pi = 0.140$, $m_q=0.140$. 
 (c) $m_\pi = 0.140$, $m_q=0.077$. 
 (d) $m_\pi = 0.140$, $m_q=0.0707$. 
Note the change in scales in the latter two panels.
The lines have the same meaning as in Fig.~\ref{figlfps.01}.
 \label{figlfss.04}}
\end{center}
\end{figure}

In Fig.~\ref{figlfss.04}, the form factors of the
scalar partner to the "pion", which we call "s-pion" in the following,
with $m_q = 0.25$, 0.14, 0.077, and 0.0707 GeV are presented for
comparison with the "pion" case. The basic features of $F_{\rm val}^+$ and
$F_{\rm nv}^+$ near $Q^2=0$ are same as in the "pion" case because
$F_{\rm val}^+(0)=1$ must hold for any meson. However, as the binding gets
weaker, we find that the "s-pion" form factors $F_{\rm tot}^{+/-}$ change
sign at smaller $Q^2$-values. This indicates that electron
scattering off the "s-pion" not only has zero cross section at a
certain electron energy, but also that the electron energy that yields zero
scattering gets smaller as the binding of the "s-pion" is weaker.
Another dramatic difference from the pseudoscalar meson is the
astonishing cancellation between $F_{\rm val}^-$ and $F_{\rm nv}^-$. Especially
in the strong binding case, both $F_{\rm val}^-$ and $F_{\rm nv}^-$ are huge but
they cancel in a very remarkable way to yield exactly the same result
as $F_{\rm tot}^+$.

\begin{figure}
\begin{center}
\hspace{17mm} (a) \hspace{46mm} (b) \hspace{46mm} (c)\\
\epsfig{figure=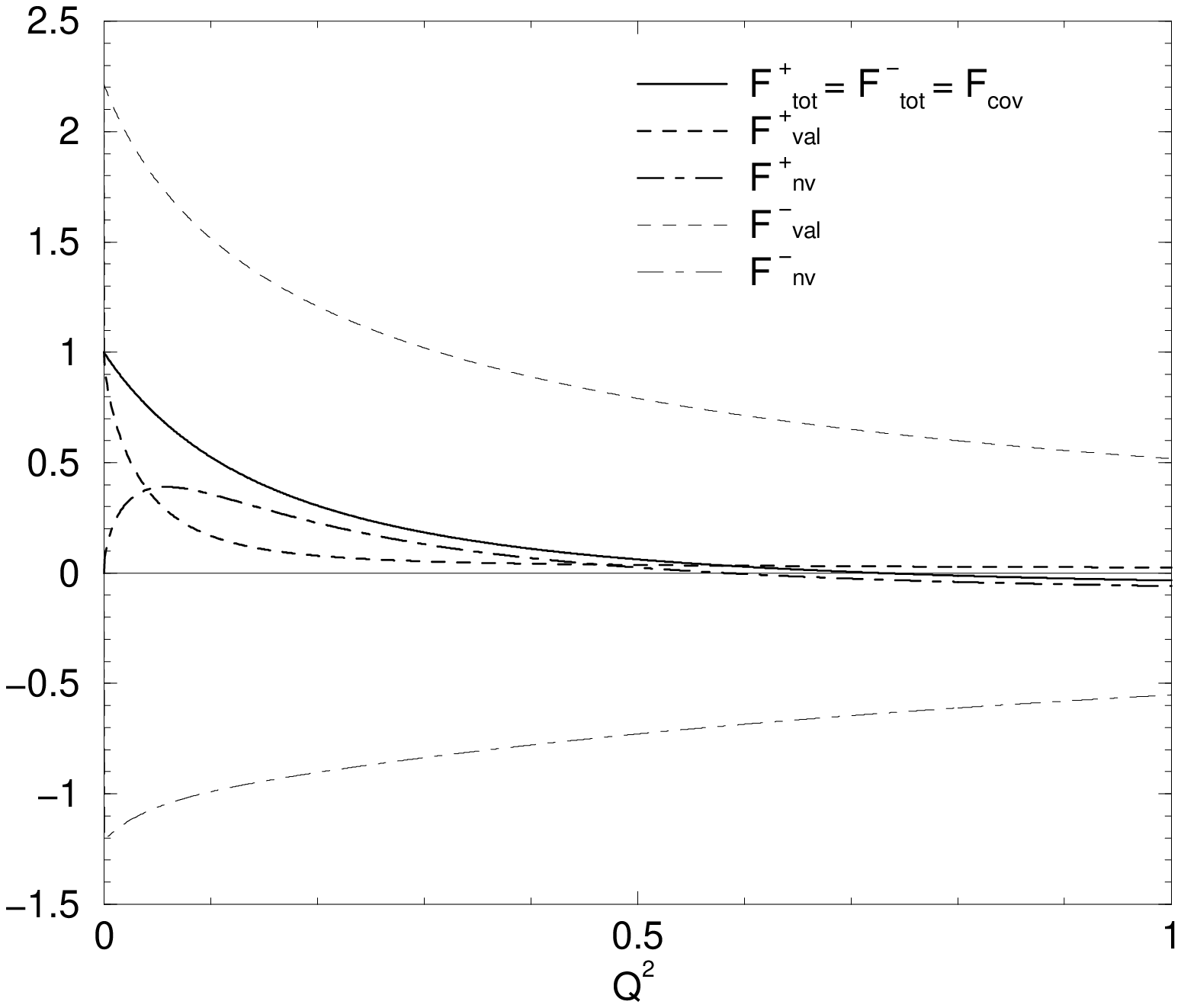,height=55mm,width=50mm} \hspace{2mm}
\epsfig{figure=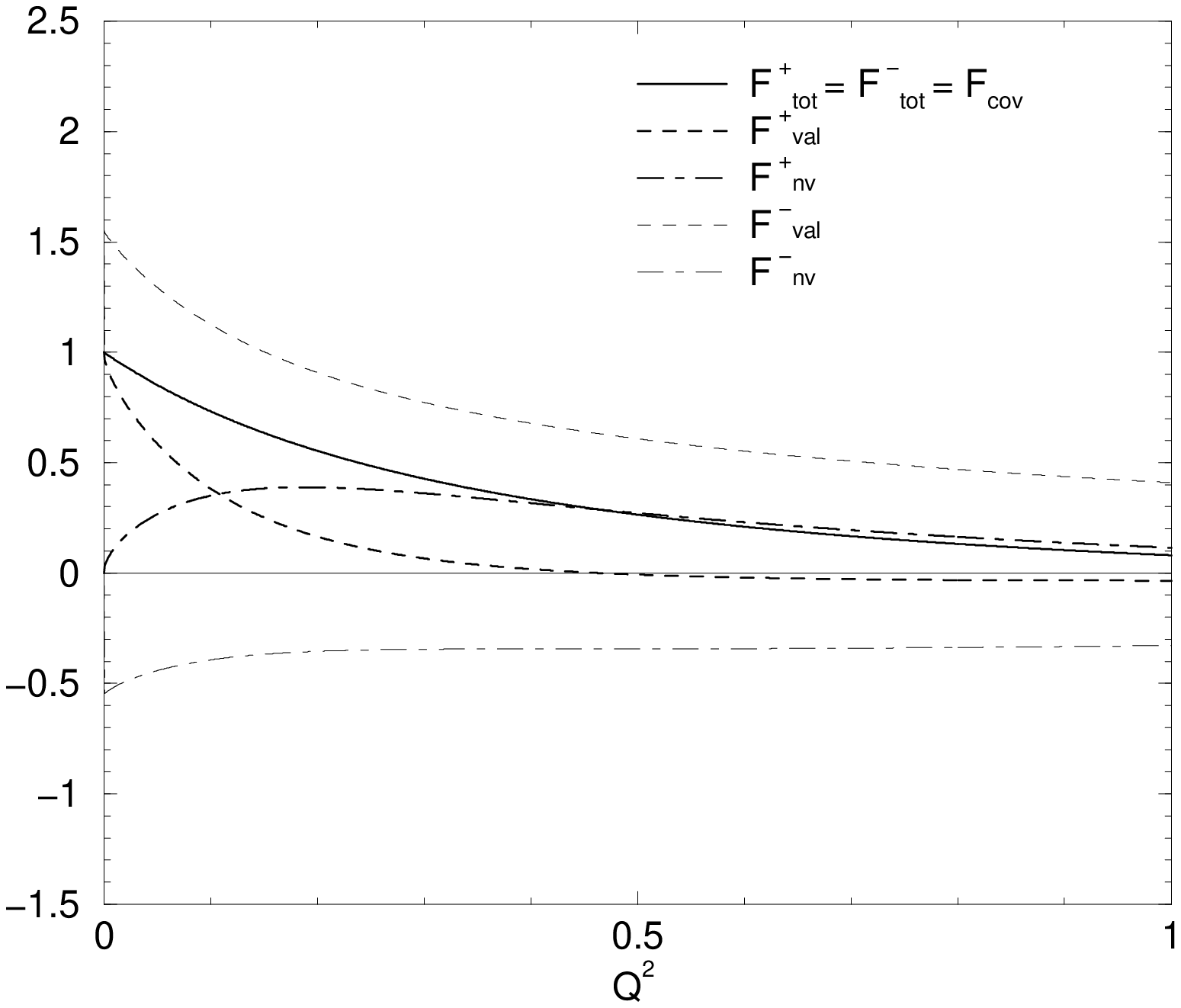,height=55mm,width=50mm} \hspace{2mm}
\epsfig{figure=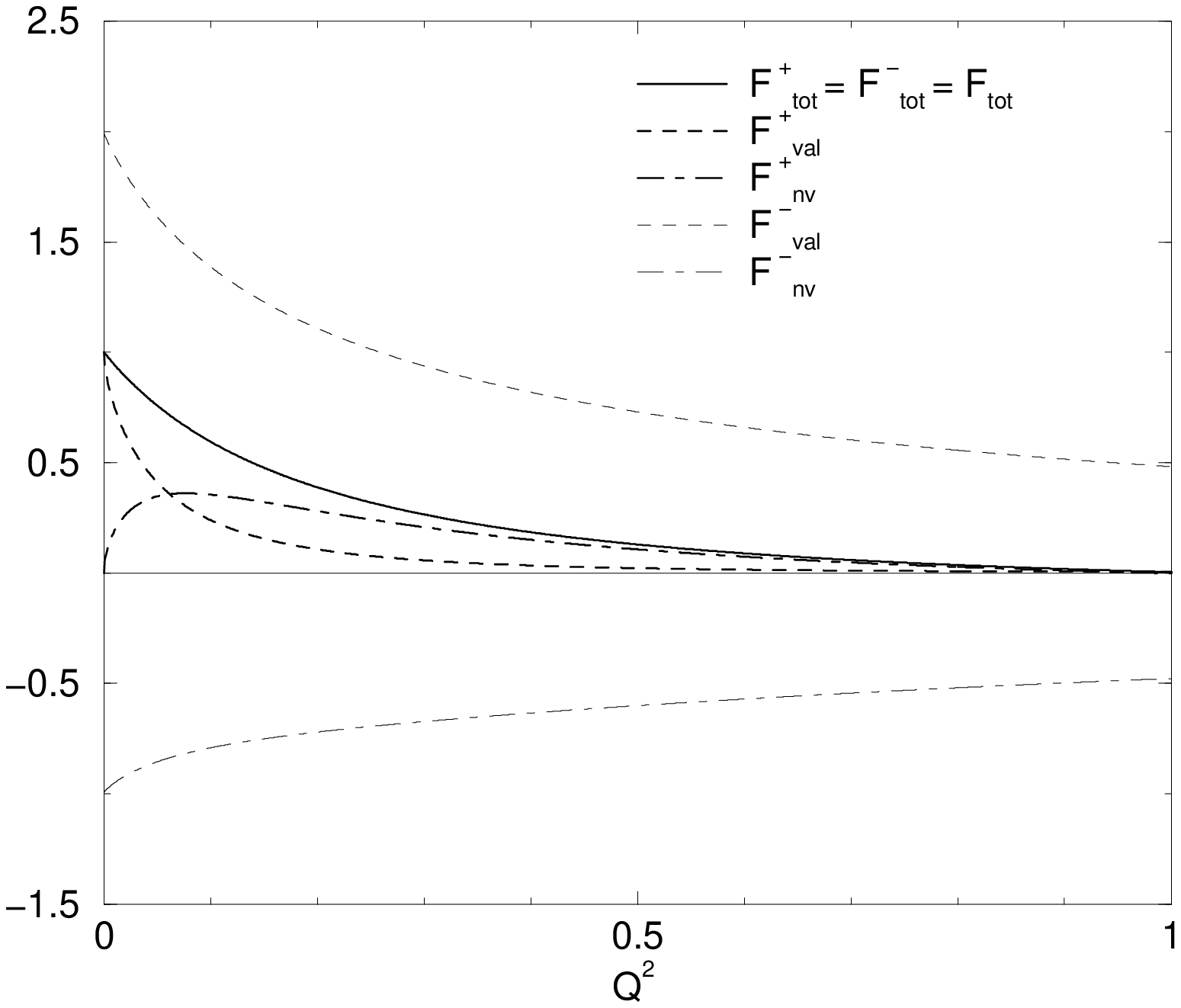,height=55mm,width=50mm}
\caption{"S-Kaon" form factor LFD calculation in 1+1 Dim. Spinor quarks.
 $m_K = 0.494$, $m_q=0.250$, $m_s=0,370$.
 (a) Charge $+1$ on the light quark, (b) charge $+1$ on the heavy quark,
 (c) $e_q = 2/3$, $e_s = 1/2$.
The lines have the same meaning as in Fig.~\ref{figlfps.01}.
 \label{figlfss.05}}
\end{center}
\end{figure}

\begin{figure}
\begin{center}
\hspace{3mm} (a) \hspace{59mm} (b) \\
\epsfig{figure=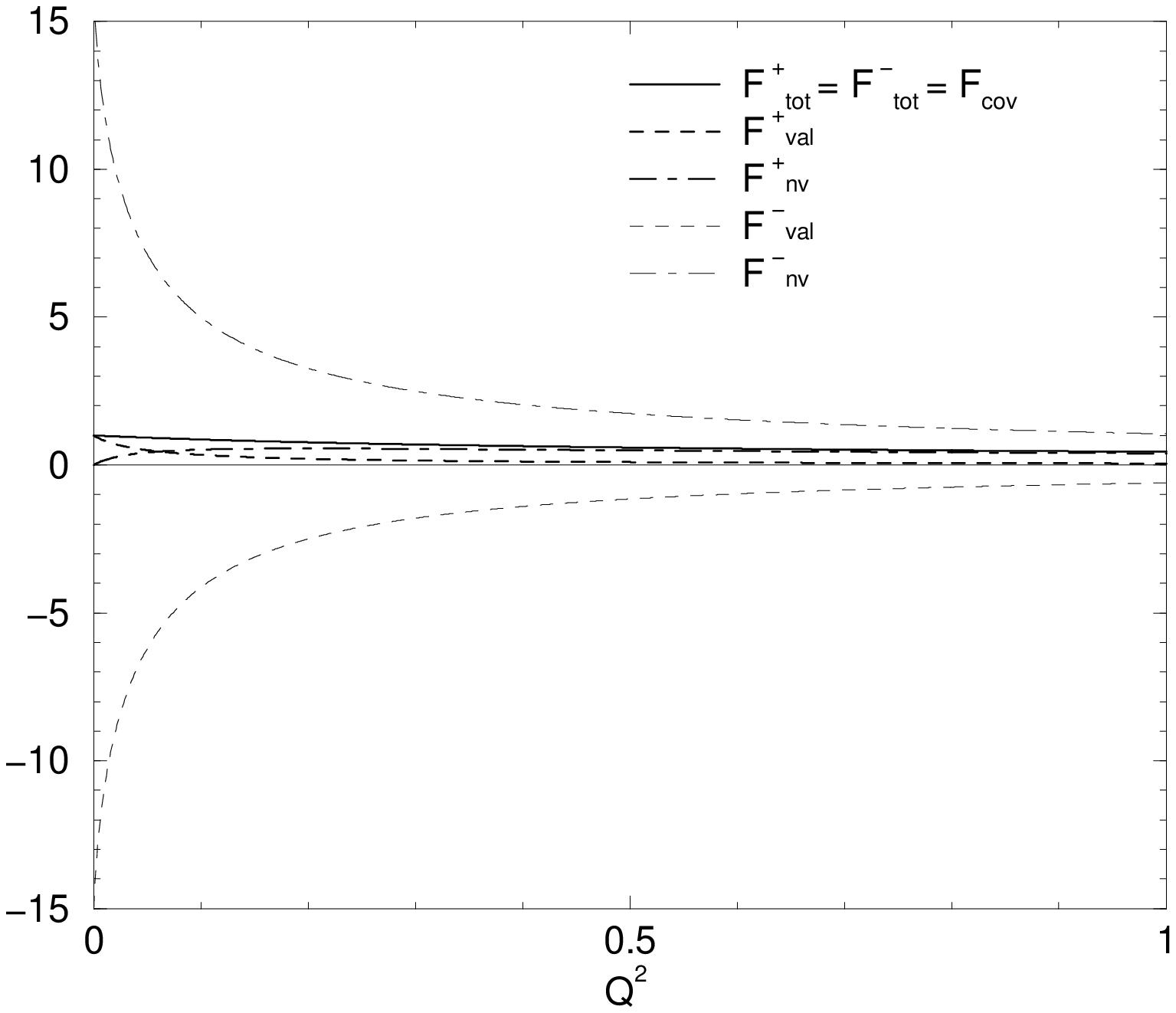,height=55mm,width=50mm} \hspace{2mm}
\epsfig{figure=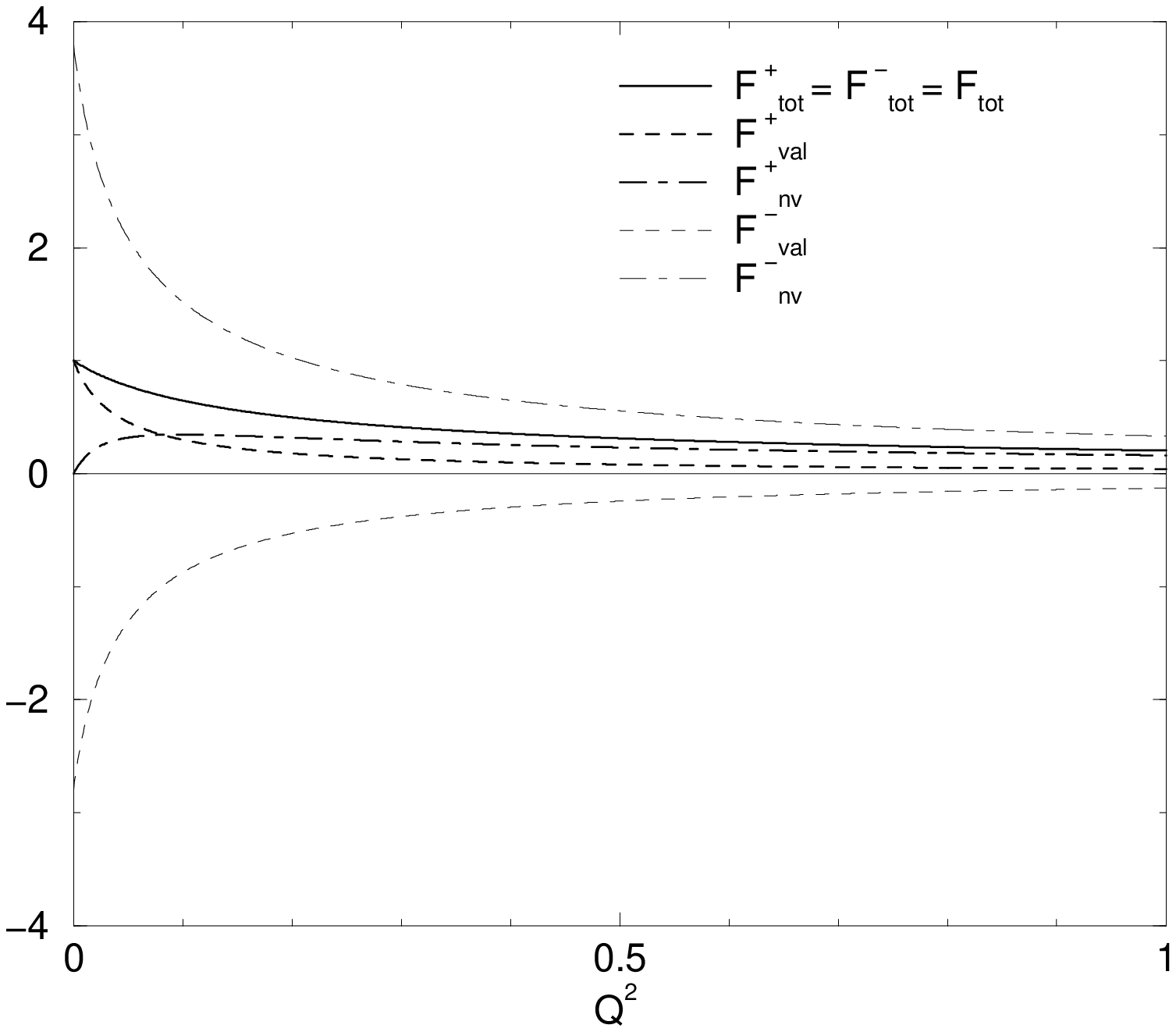,height=55mm,width=50mm} \hspace{2mm}
\end{center}
\begin{center}
\hspace{3mm} (c) \hspace{59mm} (d) \\
\epsfig{figure=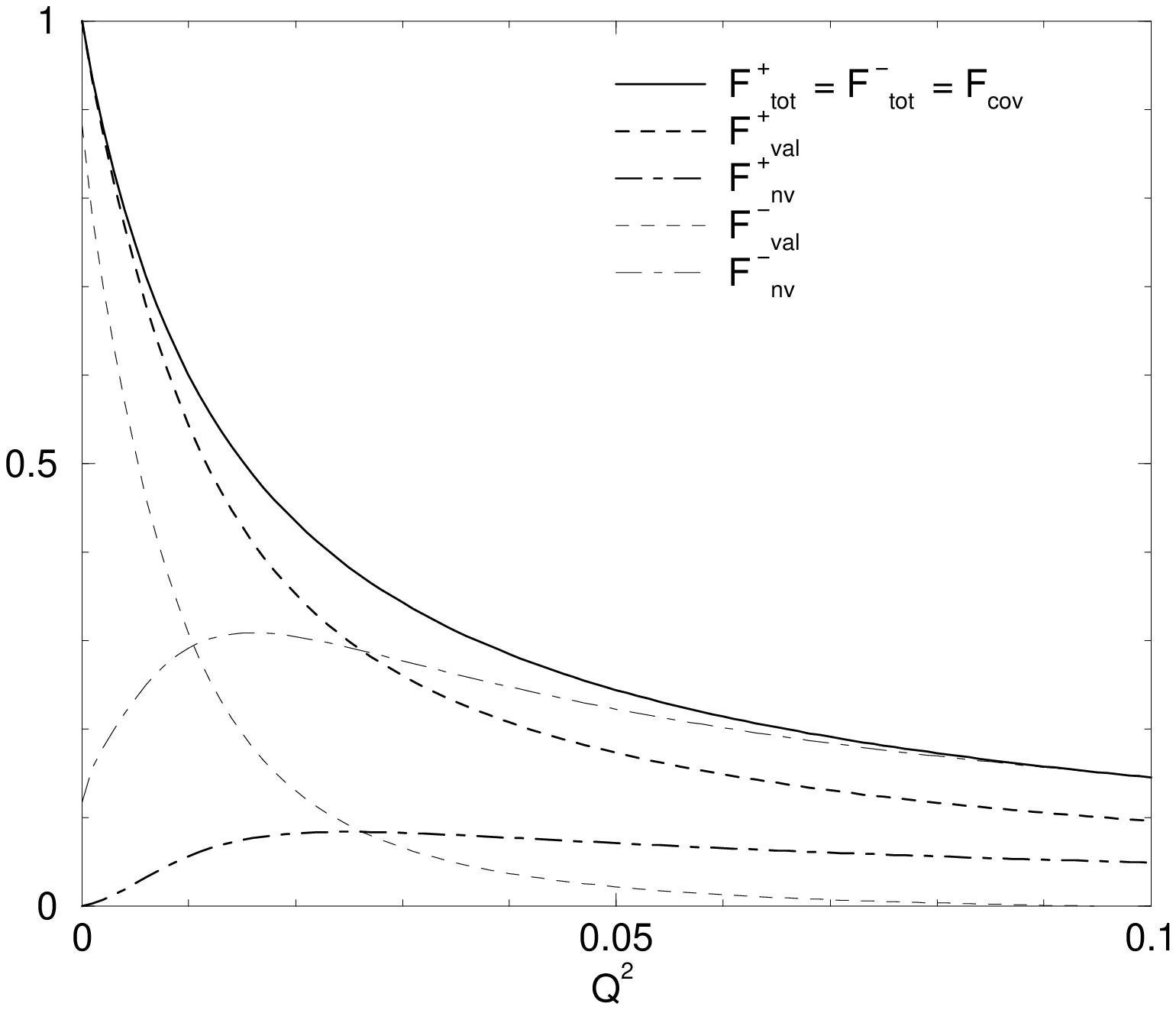,height=55mm,width=50mm} \hspace{2mm}
\epsfig{figure=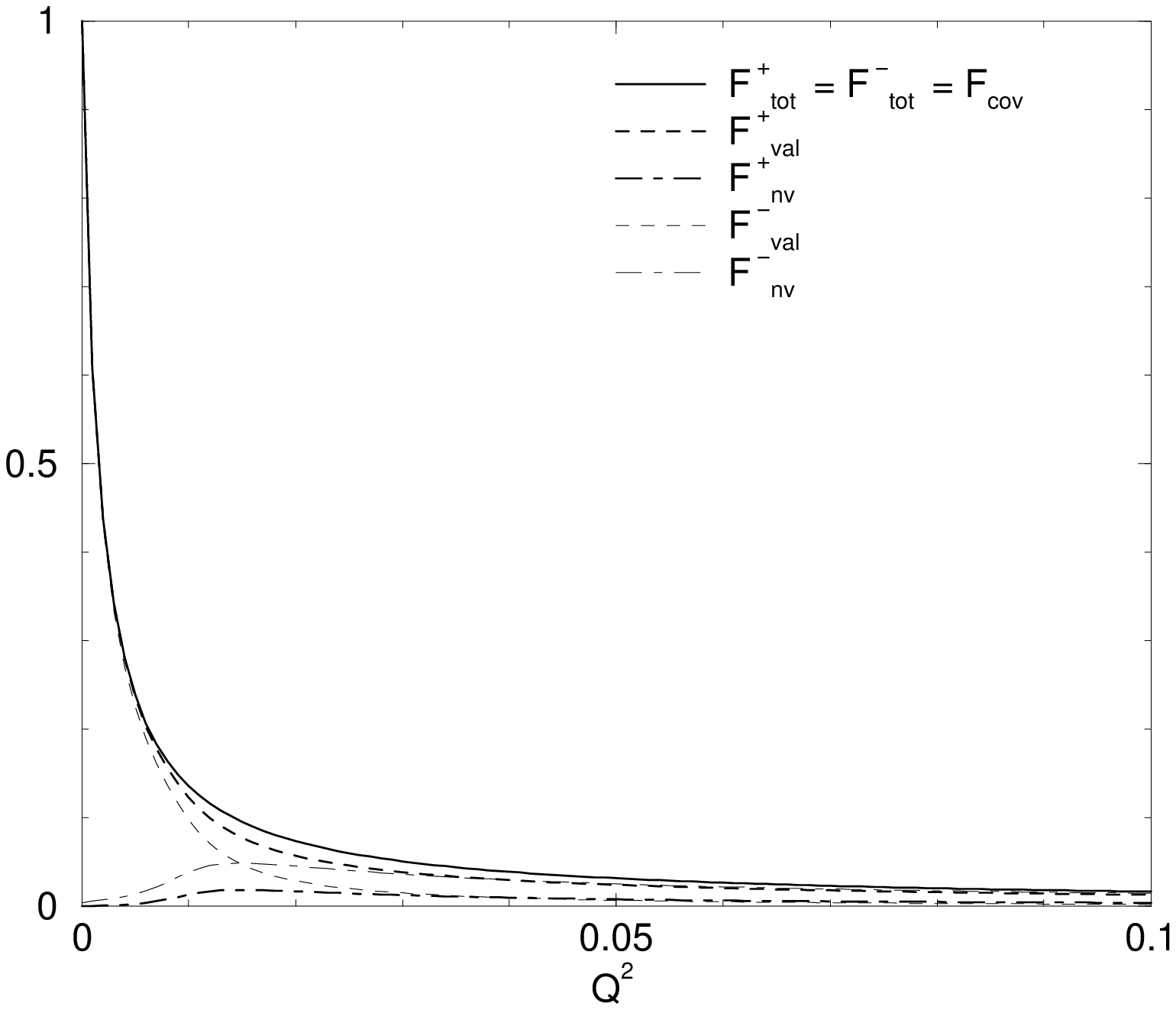,height=55mm,width=50mm}
\caption{"Pion" form factor in LF calculation in 1+1 Dim.
 Scalar meson with boson constituents.
 (a) $m_\pi = 0.140$, $m_q=0.250$. 
 (b) $m_\pi = 0.140$, $m_q=0.140$. 
 (c) $m_\pi = 0.140$, $m_q=0.077$. 
 (d) $m_\pi = 0.140$, $m_q=0.0707$. 
Note the change in scales in the latter two panels.
The lines have the same meaning as in Fig.~\ref{figlfps.01}.
 \label{figlfsb.03}}
\end{center}
\end{figure}

In Fig.~\ref{figlfss.05}, the form factor of the scalar partner to the
"kaon", {\it i.e.} "s-kaon", is plotted. The basic feature is similar
to the "s-pion".  In Fig.~\ref{figlfsb.03}, we show the results for the
"s-pion" when the spinor quark is replaced by a bosonic quark.  As we
extensively discussed in Section~\ref{sec.2}, the subtraction of the
end-point singularity is not required in the scalar quark case in
contrast to the spinor quark case. In the scalar quark case, it is
interesting to note that $F_{\rm val}^-$ and $F_{\rm nv}^-$ reveal a
large difference compare to the spinor quark case, while $F_{\rm
val}^+$ and $F_{\rm nv}^+$ are very similar to the spinor
quark case.  We find that the huge cancellation between $F_{\rm val}^-$
and $F_{\rm nv}^-$ observed in the spinor quark case does occur in
the scalar quark case only for very strong binding, and the most of
$F_{\rm tot}^-$ is saturated by $F_{\rm nv}^-$.  However, the tiny
contribution from $F_{\rm val}^-$ near $Q^2 =0$ grows as the binding
gets weaker and we have demonstrated the dominance of the valence part in
the small binding limit regardless of the spin content, as discussed
above. Fig. 9 shows the corresponding results for the "s-kaon" when the 
spinor quark is replaced by a bosonic quark. While the general features 
are similar to the spinor quark case, the bosonic quarks are more tightly 
bound together than the spinor quarks so that the charge radius of the 
meson is smaller than the case of spinor quarks as one might expect from 
the Pauli's exclusion principle. 

\begin{figure}
\begin{center}
\hspace{17mm} (a) \hspace{46mm} (b) \hspace{46mm} (c)\\
\epsfig{figure=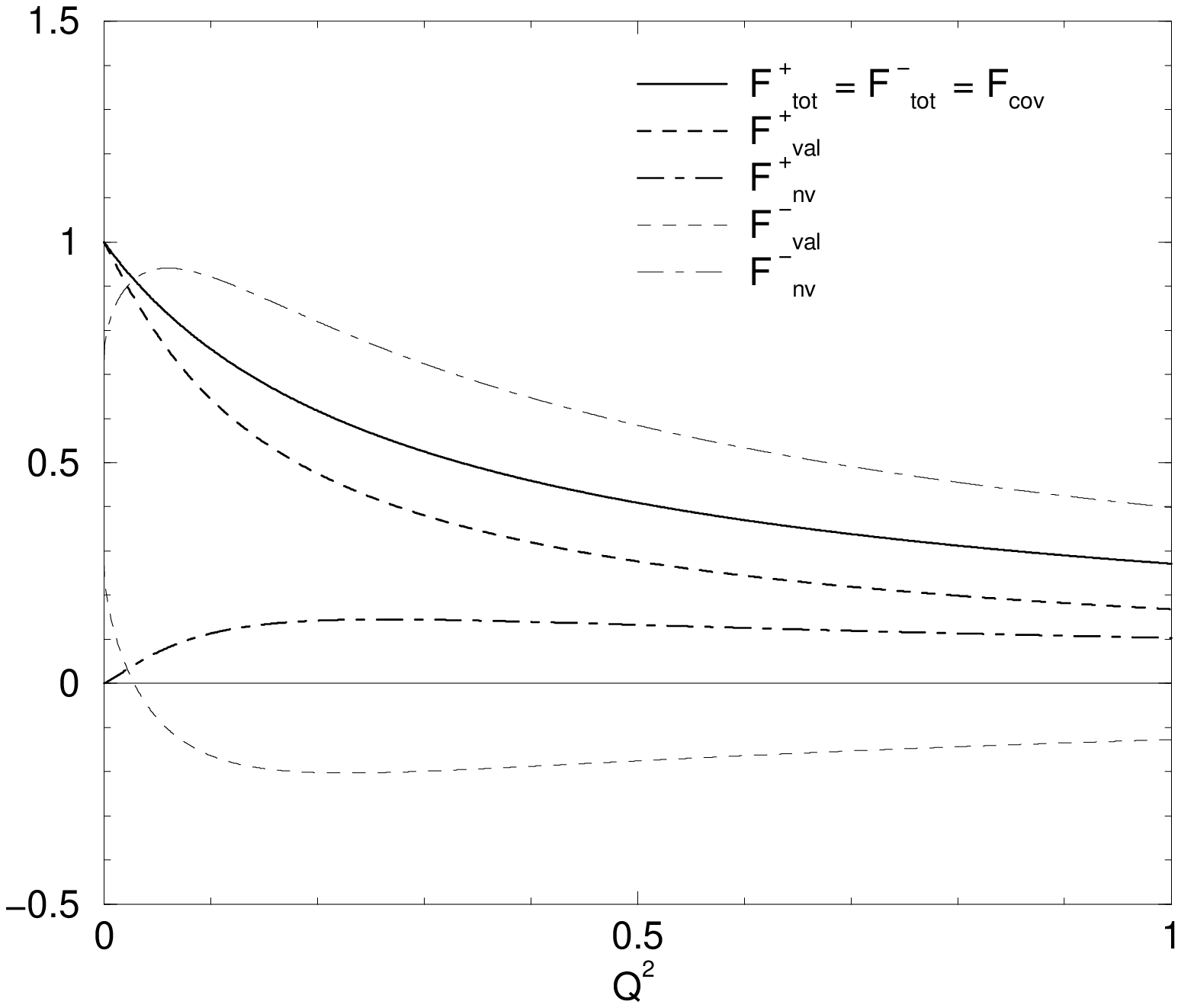,height=55mm,width=50mm} \hspace{2mm}
\epsfig{figure=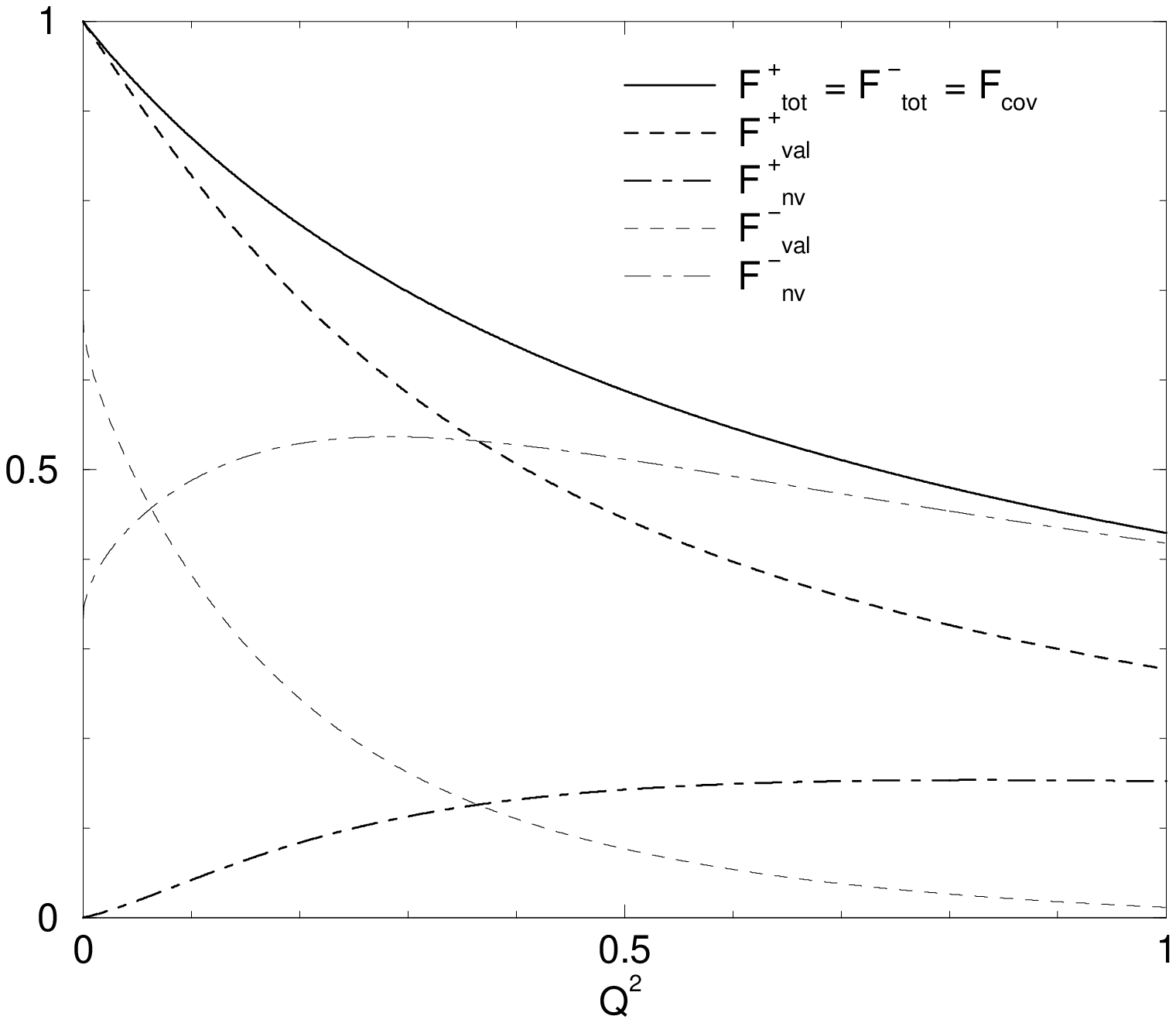,height=55mm,width=50mm} \hspace{2mm}
\epsfig{figure=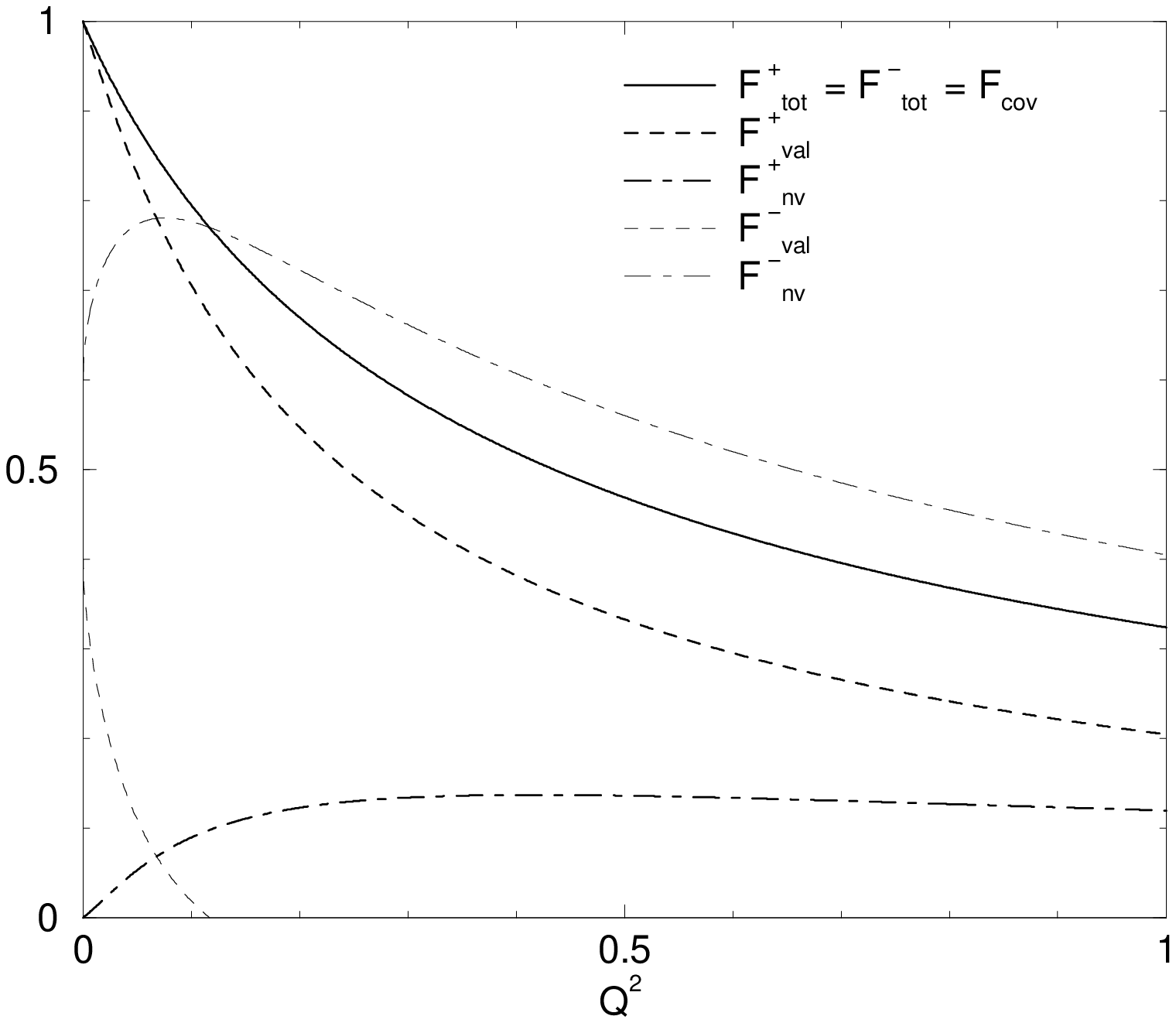,height=55mm,width=50mm}
\caption{"S-kaon" form factor LFD calculation in 1+1 Dim. Boson quarks
 $m_K = 0.494$, $m_q=0.250$, $m_s=0,370$.
 (a) Charge $+1$ on the light quark, (b) charge $+1$ on the heavy quark,
 (c) $e_q = 2/3$, $e_s = 1/2$.
The lines have the same meaning as in Fig.~\ref{figlfps.01}.
 \label{figlfsb.04}}
\end{center}
\end{figure}

\newpage
\section{Conclusion and Discussion}
\label{sec.7}

In this paper, we have analyzed both the plus and minus components of
the current quantized on the light-front to compute the electromagnetic
form factors of pseudoscalar and scalar mesons. We considered spin-1/2
consituents as well as spin-0 constituents and found dramatic
differences between the two cases. Comparing with the covariant Feynman
calculations, we notice that the common belief of equivalence between
the manifestly covariant calculation and the LF calculation linked by
the LF energy integration of the Feynman amplitude is not always
realized. The minus component of the LF current generated by the
fermion loop has a persistent end-point singularity that must be
removed to assure covariance and current conservation.  A similar
singularity was observed in the calculation of the fermion self-energy
in \cite{SB} and \cite{RM}.  The plus component of the LF current, 
however, is immune to this disorder and provides a form factor identical 
to the one
obtained doing the covariant Feynman calculation.  This phenomenon is
also associated with the spin-effect of the constituents because the
calculation with the scalar(spin-0) constituents does not have the same
symptom. Decomposing the LF amplitude into the wave-function and
non-wave-function parts, it is interesting to note that the end-point
singularity exists only in the non-wave-function vertex contribution.

\samepage
Even after the singularity is removed, the minus component of the current
sustains the zero-mode contribution while the plus component is free from 
the zero-mode. We have numerically estimated the importance of the 
non-wave-function vertices in all three cases that we discussed in 
Section~\ref{sec.2}. We considered also the unequal constituent mass 
cases such as the kaon form factor. 
We find that the behaviors of $F_{val}^-$ and $F_{nv}^-$
are trimendously different between pseudoscalar and scalar meson cases,
while $F_{val}^+$ and $F_{nv}^+$ have very similar features in both cases.
The huge but remarkably exact cancellation between $F_{val}^-$ and
$F_{nv}^-$ shown in the scalar meson case persists even if the spinor
quark is replaced by the bosonic quark. In the bosonic quark case, however,
the binding between the constituents is stronger than the spinor quark
case. We also notice that the zero-mode $F_{nv}^-(0)$ diminishes as the 
binding gets
weaker. Our results are quite consistent to the earlier observation
\cite{CJ} exhibiting the smaller zero-mode contribution in the heavier
quark systems. In all of these cases, our results 
show that if the meson is weakly bound then the contributions from the 
wave-function and the non-wave-function vertices to the plus current are 
separately almost the same as those for the minus current. Of course, 
their sums add up to the same number as the covariant Feynman result 
in both the plus and minus cases.

The calculations carried out so far are semi-realistic as the model was
1+1-dimensional and only a point-vertex was considered.  It is clear
from a formal analysis of the 3+1-dimensional case, however, that a
singularity of the same form will occur in the matrix element of $J^-$
calculated in LFD regardless of dimensionality.  A recent analysis of the
Burkhardt-Cottingham sum-rule seems to reveal a similar divergence in the
polarized spin-1/2 structure functions \cite{Matthias}.  While the additional
regularization may be provided by smearing the point-vertex with a
realistic wave-function in the 3+1-dimensional covariant treatment of
the current, the identification of the singular term as we achieved in
this work would still be necessary for the smeared vertex cases. The
importance of the non-wave-function parts may nevertheless differ
numerically from the 1+1-dimensional case.  This point is presently
under investigation.

\acknowledgements
\noindent
This work was supported in part by a grant from the US Department of
Energy and the Netherlands Organisation for Scientific Research (NWO).

\end{document}